\documentstyle[aaspp4]{article}
\input epsf
\begin{document}
\title{Supersonic motions in dark clouds are {\it not} Alfv\'{e}n waves}

\author{Paolo Padoan}
\affil{Theoretical Astrophysics Center, Juliane Maries Vej 30, 
DK-2100 Copenhagen, Denmark}
\author{\AA ke Nordlund}
\affil{Astronomical Observatory and Theoretical Astrophysics Center, 
Juliane Maries Vej 30, DK-2100 Copenhagen, Denmark}
\authoremail{padoan@tac.dk}

\begin{abstract}

Supersonic random motions are observed in dark clouds and are 
traditionally interpreted as Alfv\'{e}n waves, but the possibility 
that these motions are super-Alfv\'{e}nic has not been ruled out. 

In this work we report the results of numerical experiments in two 
opposite regimes; $\beta\approx 1$ and $\beta\ll 1$, where $\beta$ 
is the ratio of gas pressure and magnetic pressure: $\beta=P_{g}/P_{m}$. 

Our results, combined with observational tests, show that the model with 
$\beta\approx 1$ is consistent with the observed properties of molecular 
clouds, while the model with $\beta \ll 1$ has several properties that
are in conflict with the observations.  

We also find that both the density and the magnetic field in molecular
clouds may be very intermittent.  The statistical distributions of 
magnetic field and gas density are related by a power law, with an 
index that decreases with time. Magnetically dominated cores form 
early in the evolution, while later on the intermittency in the 
density field wins out.

\end{abstract}


\section{Introduction}

The observation of supersonic motions in molecular clouds (eg Zuckerman 
\& Palmer 1974) raised the question of how these motions could be supported 
(Norman \& Silk 1980; Fleck 1981; Scalo \& Pumphrey 1982). 
Supersonic motions are expected to quickly dissipate their energy in highly 
radiative shocks, because of the very short cooling time of molecular gas or 
metal rich atomic gas (Mestel \& Spitzer 1956; Spitzer 1968; Goldreich \& 
Kwan 1974). 

Strictly related was the issue of the support of molecular clouds (MCs) 
against gravitational collapse, since it was soon realized that the observed 
motions could not be understood as a gravitational collapse (Zuckerman \& 
Evans 1974; Morris et al.\ 1974), although MCs contain many Jeans' masses. 

Theoreticians therefore formulated the hypothesis that MCs were primarily 
magnetically supported (Mestel 1965, Strittmatter 1966; Parker 1973;
Mouschovias 1976a,b; McKee \& Zweibel 1995) and interpreted the observed 
motions as long-wavelength hydro-magnetic waves (Arons \& Max 1975; Zweibel 
\& Josafatsson 1983; Elmegreen 1985, Falgarone \& Puget 1986). It was also 
shown that the properties of the observed `turbulence' (Larson 1981; Leung, 
Kutner \& Mead 1982; Myers 1983; Quiroga 1983; Sanders, Scoville \& Solomon 
1985; Goldsmith \& Arquilla 1985; Dame et al.\ 1986; Falgarone \& P\'{e}rault 
1987) could be understood if the motions were sub-Alfv\'{e}nic (Myers \&
Goodman 1988; Mouschovias \& Psaltis 1995, Xie 1996).

Nevertheless, recent attempts to detect the Zeeman effect, in lines of 
molecules such as OH and CN, that probe regions of dense gas (Crutcher 
et al.\ 1993; Crutcher et al.\ 1996), resulted in a number of 
non-detections of the effect, and therefore in rather stringent upper 
limits for the magnetic field strength, despite the fact that regions 
expected to favor detections were targeted. It is therefore possible, 
that the field strength in dense molecular gas is weaker than assumed
in theoretical studies.

In the present work, we describe the dynamics of MCs, with the 
numerical solution of the equations of 3-D compressible 
magneto-hydrodynamics 
(MHD), in a regime of highly supersonic random motions. We avoid 
any sort of 
idealization of the physical system, apart from describing it as a 
fluid, and 
simply rely on the physical assumption (supported by the observations) 
that the 
motions in MCs are random. The main limitation of
our numerical models, compared with MCs, is the absence of gravity.
We have excluded gravity on purpose (our code is capable of handling
self-gravity), because one of the aim of our work
is to show that (magneto-)hydrodynamic processes alone are able to 
explain most
of the observed properties of MCs, without the need of gravity.
Although gravity is certainly responsible for the fragmentation into stars
of high density regions, supersonic random motions shape the structure of
molecular clouds with a minor contribution from gravity (Padoan 1995; 
Padoan, Jones \& Nordlund 1997a; Padoan, Nordlund \& Jones 1997b), apart 
from the
possibility that gravity is the energy source of the motions on the 
large scale.
The importance of supersonic motions in fragmenting the gas is certainly 
apparent
on very small scales, where young and probably transient clumps are 
not originated by gravitational instability (Falgarone, Puget \& 
P\'{e}rault 1992; 
Langer et al. 1995).

Previous numerical studies have shown that compressible turbulence can
qualitatively explain several observational properties of MCs, even if
the effect of gravity and magnetic fields are not included.
The first two-dimensional (2-D) simulations of turbulence, with rms 
Mach number
larger than one, were performed by Passot \& Pouquet (1987). 
These were the
first simulations where shocks are shown to develop inside a turbulent 
flow. The importance of shock formation inside mildly supersonic flows
was later confirmed in 3-D simulations by Lee, Lele \& Moin (1991). It
was immediately recognized that shocks might have been responsible
for the fragmentation of the density field inside MCs, and especially
for the origin of their filamentary structure (Passot, Pouquet \& 
Woodward 1988).
Kimura \& Tosa (1993) simulated the passage of a strong shock through a
turbulent molecular cloud, and found that this process can generate dense
clumps, with a power law mass spectrum. V\'{a}zquez-Semadeni (1994) made
use of 2-D numerical simulations to show that supersonic turbulence
generates a very intermittent density field,
reminiscent of the clumpy nature of MCs. The density field was also found 
to be self-similar, which could be the reason for the hierarchical structure
of MCs (Scalo 1985, Houlahan \& Scalo 1992). Falgarone et al. (1994), 
analyzing 
the 
numerical simulation by Porter, Pouquet \& Woodward (1994), argued 
that the properties of the profiles of molecular emission spectra 
from MCs can be interpreted as arising from turbulent motions. 

Other numerical works included gravity in the turbulent flows, yet 
without describing the magnetic field. Turbulence was shown to be able
to prevent gravitational collapse (cf Chandrasekhar 1958; Arny 1971; 
Bonazzola et al. 1987, 1992) in the 2-D numerical simulations by 
L\'{e}orat, Passot \& Pouquet (1990). V\'{a}zquez-Semadeni, Passot \&
Pouquet (1995) modeled the galactic disc, on the scale of 1 Kpc, as a 
turbulent self-gravitating flow. They simulate a 2-D turbulent flow
that is forced
by the energy released by star formation (expansions of HII regions), 
and found
that the main mechanism of cloud formation is the turbulent ram pressure,
rather than gravity. They were not able to form self gravitating clouds, 
due
to limitations in the thermal modeling, and the consequent low density 
contrast.

A 3-D description of a magnetized self-gravitating cloud was
given by Carlberg \& Pudritz (1990), and was used to simulate 
molecular emission spectra by Stenholm \& Pudritz (1993). 
Carlberg \& Pudritz found that the magnetic field and hydro-magnetic
waves can support the cloud against gravity. The clouds contract,
because of ambipolar diffusion, on a timescale of approximately four 
free-fall times. These simulations do not solve the MHD equations, but
instead make use of a 'sticky particles' code. Energy is injected
in the form of a spectrum of Alfv\'{e}n waves, and the outcome of the
computation is dependent on the spectral index, that is a free 
parameter.
This way of forcing the particles is rather unphysical, 
because an arbitrary spectrum of Alfv\'{e}n waves is imposed, 
instead of being obtained as a result of the simulated 
magneto-hydrodynamics. Passot, V\'{a}zquez-Semadeni \&
Pouquet (1995) introduced the magnetic field in their previous 2-D model
for the galactic disc (V\'{a}zquez-Semadeni, Passot \& Pouquet 1995), 
and obtained a flow with rough equipartition of kinetic and magnetic energy,
probably in rough equipartition also with the mean thermal energy.
The same simulation, and others with larger density contrast and resolution,
have been studied by V\'{a}zquez-Semadeni, Ballesteros-Paredes \& 
Rodr\'{i}guez (1997), who were able to reproduce the observed 
relation between line-width and size (Larson 1981). Gammie and 
Ostriker (1996) solved the MHD equations in a slab geometry, 
including self-gravity. By forcing the flow with
a nonlinear spectrum of MHD waves, they were able to prevent the 
gravitational collapse of their 1-D cloud model.

Apart from the intentional exclusion of gravity, we have improved
significantly on all previous calculations of turbulent flows. First of
all we have solved the MHD equations in three-dimensions, while all
previous solutions are in two or one dimensions. In MHD the dimensionality
of the flow has a fundamental importance, because it determines the
topological freedom of the magnetic field. Another improvement of
our work is the high rms Mach number of the flows ($\sim$ 5), while previous 
models
are only mildly supersonic (Mach number $\sim$ 1). While previous models of
magnetized clouds focused on the role of the magnetic field as opposed to
gravity, and therefore have assumed a magnetic pressure much larger than the
gas pressure, the main purpose of the present work is to show that MCs are 
well described as flows with much lower magnetic pressure than previously 
assumed, and probably in rough equipartition with the gas pressure.

We report on the results of two numerical models. In model A, 
$\beta\approx 1$, in model B, $\beta\ll 1$, where $\beta$ is the ratio of gas 
pressure and magnetic pressure: $\beta=P_{g}/P_{m}$. The observations 
of magnetic field strengths are in agreement with a scenario where the mean 
magnetic pressure, $\langle B^2a\rangle$, is dynamically low (model A). 
Magneto-hydrodynamic (MHD) flows develop a very intermittent 
spatial distribution of the magnetic energy, and therefore, when the 
field is detected at a favorable position, its strength is far stronger 
than the mean field strength. 
Dense cores with sub-Alfv\'{e}nic velocity dispersion can still be generated, 
in agreement with the observations.

\section{The experiments}

The study of the dynamics of MCs belongs to the field of random MHD 
flows, in a highly supersonic regime. The Reynolds number and the magnetic 
Reynolds number in MC flows are very large. Their random nature is therefore
a basic feature of the dynamics of these flows, and requires an appropriate 
description.

For this reason, a realistic description of the dynamics
of molecular clouds had to be based on the numerical solution of
the compressible MHD equations in three dimensions, in a random and
highly supersonic regime. We did it with relatively high resolution
($128^3$), with a code designed for turbulence and MHD turbulence experiments
(Nordlund, Galsgaard \& Stein 1994; Stein, Galsgaard \& Nordlund 1994; 
Galsgaard \& Nordlund 1996; Nordlund, Stein \& Galsgaard 1996; Nordlund \& 
Galsgaard 1997),
specifically adapted to be able to deal with very strong shocks and very large 
density contrasts. 

One of our purposes with these particular experiments is to
explicitly address the classical arguments about dissipation time scales
for super- and sub-Alfv\'{e}nic motions,  and we therefore perform the
experiments in the spirit of ``freely decaying turbulence''; studying the
decay of an initial, solenoidal velocity field, without applying any 
external forcing. We include external forcing in other experiments that we
have made (e.g., Padoan et al.\ 1997a, 1997b).

Although we have already developed a version of the code with the inclusion
of the gravitational force, all the experiments were run without gravity, 
for the following reasons:

\begin{itemize}

\item We are mainly interested in studying the magneto-hydrodynamics,
rather than the gravitational instability.

\item The observed motions have velocities comparable with the virial
velocity, or larger, on a range of scales, and the clouds are not 
free-falling.

\item If the results of our experiments are discussed only up to a time
shorter than or comparable to the dynamical  (or free-fall) time, all our
conclusions remain basically unchanged. This time is about a few million 
years on a scale of 10 pc, and clouds are not supposed to live much 
longer than that, before star formation takes place and becomes 
energetically important.

\end{itemize}

\subsection{The equations}

We solve the compressible MHD equations:

\def\vv{{\bf v}}
\def\jj{{\bf j}}
\def\bb{{\bf B}}
\def\lnr{\ln\rho}
\def\div{\nabla\cdot}

\begin{equation}
\label{0}
{\partial \ln\rho \over \partial t} + \vv \cdot \nabla\lnr = - \div \vv,
\end{equation}

  \begin{equation}
   {\partial{\vv} \over \partial t}
   + {\vv\cdot\nabla\vv}
  =
   - {1\over\rho} \nabla P
   + {1\over\rho} {\jj} \times {\bb} + {\bf f},
  \label{1}
  \end{equation}

\begin{equation}
\label{4}
{\partial e \over \partial t} + {\vv} \cdot \nabla e = - {P \over \rho} \div 
{\vv} + Q_{\rm Joule} + Q_{\rm viscous} + Q_{\rm radiative},
\end{equation}

\begin{equation}
{\partial{\bb} \over \partial t} = \nabla\times\vv\times\bb,
\label{2}
\end{equation}

\begin{equation}
\jj = \nabla\times\bb,
\label{3}
\end{equation}

\noindent
plus numerical diffusion terms, and
with periodic boundary conditions. $\vv$ is the velocity, $\bb$ the
magnetic field, ${\bf f}$ an external force ($=0$ in these particular
experiments), and $p = \rho T$ is the pressure at $T \approx$ const. 

Conditions in the cold molecular clouds that we are modeling are such that
an isothermal approximation is adequate; the radiative heat exchange is
so efficient that the temperature remains low in most places.  Even if
the temperature momentarily increases in shocks, the subsequent cooling is
rapid, and the result is shock structures that are qualitatively and
quantitatively similar to isothermal shocks.

We have thus used isothermal conditions in most of our runs and
have verified that this is appropriate, by
rerunning segments of some experiments using the full energy equation.
No significant change of the statistics was found and, since using the
full energy equation increases the cost of the experiments considerably
(the strong cooling required to maintain a low temperature forces a
much smaller time step), we performed most of the experiments at
constant temperature.

The absence of an explicit resistivity $\eta$ in the induction equation 
corresponds to an assumption of flux freezing on well resolved scales.
The code uses shock and current sheet capturing techniques to ensure
that magnetic and viscous dissipation at the smallest resolved scales 
provide the necessary dissipation paths for magnetic and kinetic energy.
As shown by Galsgaard \& Nordlund (1976, 1977), dissipation of 
magnetic energy in highly turbulent MHD plasmas occurs at a rate that
is independent of the details of the small scale dissipation. 
In ordinary hydrodynamic turbulence the corresponding property is
one of the cornerstones of Kolmogorov (1941) scaling.

We have not included ambipolar diffusion in the present experiments,
because it occurs on a time-scale significantly longer than the 
dynamical time, as recently shown by Myers \& Khersonsky (1995) and
is thus expected to be of secondary importance.  However, our 
code has the ability to handle ambipolar diffusion, and we plan 
to study its effects in a subsequent paper.

\subsection{The code}

The code solves the compressible MHD equations on a 3D staggered mesh, 
with volume centered mass density and thermal energy, face centered 
velocity and magnetic field components, and edge centered electric 
currents and electric fields (Nordlund, Stein \& Galsgaard 1996).

The original code works with ``per-unit-volume'' variables; mass density,
momenta, and thermal energy per unit volume.  In the super-sonic regime 
relevant in the present application, we found it advantageous to 
rewrite the code in terms of ``per-unit-mass'' variables; $ln\rho$, $u$,
and $E=\frac{3}{2}\frac{P}{\rho}$.
With these variables, the time evolution of all variables is
governed by equations of the type
\begin{equation}
{D f \over Dt} = {\partial f \over \partial t} + \vv \cdot f = ... ;
\end{equation}
i.e., equations that specify the time rate of change following the motion.
These are better conditioned than the divergence type equations that
result from using per-unit-volume variables (the large---order $M^2$---
density variation in isothermal shocks cause the per-unit-volume fluxes to
vary over several orders of magnitude).

We use spatial derivatives accurate to
6th order, interpolation accurate to 5th order, and Hyman's 3rd order time 
stepping
method (Hyman 1979).

In order to minimize the viscous and resistive influence on well resolved
scales, we use monotonic 3rd order hyper-diffusive fluxes instead of normal
diffusive fluxes, and in order to capture hydrodynamic and
magneto-hydrodynamic shocks we add diffusivities proportional to the
negative part of the velocity divergence, and resistivity proportional to
the negative part of the cross-field (two-dimensional) velocity divergence.
Further details of the numerical methods are given by Nordlund, 
Galsgaard \& Stein (1996) and Nordlund \& Galsgaard (1997).

\subsection{Weak and strong magnetic field}

For the purpose of this work we have run two experiments: 
one with $\beta_{i}=4$ (model A), and the other with $\beta_{i}=0.04$
(model B), where $\beta_{i}=(P_{g}/P_{m})_{i}$ is the initial ratio
of gas and magnetic pressure.

In both experiments the initial density is uniform, and the initial
velocity is random. We generate the velocity field in Fourier space,
and we give power, with a normal distribution, only to the Fourier
components in the shell of wave-numbers $1<k<2$. We perform
a Helmholtz decomposition, and use only the solenoidal 
component of the initial velocity. The initial magnetic field is 
uniform, and is oriented parallel to the $z$ axis: 
${\mathbf B}=B_0{\mathbf \hat{z}}$. 

In experiment B, the initial velocity field is perpendicular
to the magnetic field (zero $z$ component), in order to excite 
the Alfv\'{e}n waves, while in experiment A the initial 
velocity field has equal amplitude in all three components.  
Both experiments are decaying flows, because no 
external forcing is applied. 

We define the Alfv\'{e}nic Mach number, ${\cal{M}}_{ A}$, as 
the ratio of the flow rms velocity and the Alfv\'{e}n velocity. 
In model A, ${{\cal{M}}_{ A}}\approx 10$, and in model B,
${\cal{M}_{ A}}\approx 1$, initially.  
Under isothermal conditions, in our units, the ordinary Mach number 
${\cal{M}} = {\cal{M}}_{ A} / \beta^{1/2}$, so the ordinary 
Mach number is initially ${\cal{M}} \approx 5$ in both cases.

Fig.~1 shows the time evolution of the rms Mach number and 
rms density in the two experiments. In both cases, the
rms density grows to the value $\sigma_{n}\approx 0.5 {\cal{M}}$, 
where ${\cal{M}}$ is the rms Mach number of the flow, in about one 
dynamical time, defined as the ratio of half the linear size of 
the box over the initial rms velocity, 
$t_{dyn}=0.5L_{box}/\sigma_{v,0}$. 

The initial Mach number in experiment B is a bit smaller than 
in experiment A, because the amplitude of the initial $z$ 
component of the velocity field is zero. Nevertheless, 
energy is immediately transferred to the $z$ component 
of the velocity, and the rms $z$ component is subsequently about half 
the value of the other two components.

Fig.~2 illustrates the time evolution of the kinetic and magnetic 
energies, expressed in units of the mean thermal energy. The 
time evolution of the magnetic energy is totally different 
in the two experiments. In model B the field lines are only 
weakly perturbed by the flow, and exchange their energy with the 
flow periodically (oscillations in magnetic and kinetic energy, 
$E_m$ and $E_k$); in model A, instead, the magnetic field is 
almost passively advected, and its intensity and direction 
are strongly influenced by compressions and stretching 
of field lines. $E_m$ grows until it is in approximate 
equipartition with the thermal energy of the gas, but 
remains well below equipartition with the kinetic energy.

\section{Observational tests}

In this section we compare our numerical results with
observed properties of molecular clouds.

\subsection{Dissipation of supersonic motions}

The time evolution of the kinetic energy is plotted in Fig.~2.
After two dynamical times, $t=2.0t_{dyn}$, the flow
in experiment A still contains about $30\%$ of its initial 
kinetic energy or, in other words, the rms velocity is still 
about $55\%$ of its initial value. In experiment B, the mean 
kinetic energy of the oscillations after two dynamical times 
is about $40\%$ of its initial value, i.e.\ the rms
velocity is about $63\%$ of the initial value.  

Thus, contrary to the beliefs that partly motivated 
developing the Alfv\'{e}n wave model of MC turbulence, 
the dissipation in model A is not particularly rapid in absolute 
terms, and the dissipation in model B is not particularly small.  
Thus, the advantage of the Alfv\'{e}n wave model,
as far as the energy dissipation time-scale is concerned, is little: 
$t_{diss,B}=1.3t_{diss,A}$. Although this was already appreciated 
in Zweibel \& Josafatsson (1983) and Elmegreen (1985), it should 
be noticed that here we do not include the dissipation mechanism 
of ion-neutral friction. Alfv\'{e}n waves dissipate mainly 
because they transfer their energy to motions along the magnetic 
field, that are eventually dissipated in shocks. This was found 
also by Gammie and Ostriker (1996). It is clear therefore 
that, from the point of view of the dissipation 
mechanism, model B is not significantly superior to model A.

The dissipation of the supersonic motions is not a problem 
for any of the models, however, because:

\begin{itemize}

\item If motions are present there must be a source for their energy;
such a source may be active for more than one dynamical time.

\item After one dynamical time $50\%$ of the initial energy 
in model A is still in the flow. At the scale of $10$ pc the 
dynamical time is about 
$3\times10^6$ years: clouds do not need to exist much longer than that 
before star formation significantly affects their energetics 
(Blitz \& Shu 1980). 

\end{itemize}

\subsection{Magnetic energy probability distribution}

We illustrate the probability distribution of the magnetic energy 
in units of the mean thermal pressure: $1/\beta=B^2/<P_g\rangle$. 
We recall that initially
$\beta$ is uniform in both models, and its value is $\beta_{i}=4$ 
in model A, and  $\beta_{i}=0.04$ in model B.

The distribution of the magnetic energy is shown in Fig.~3.
Model A develops a very intermittent distribution, with an 
exponential tail.

Taking into account the relation between field strength and density
(see next section), one finds that approximately $0.5\%$ of the total 
mass of the system contains a field 10 times stronger than the mean value. 
For example, in a cloud of $10^{4}M_{\odot}$, one can find a couple 
of clumps of about $30M_{\odot}$ each, with a field strength 
$B=40\mu G$, while the mean field of the cloud is only $B=4\mu G$.

Model B is much less intermittent than model A.

\subsection{The relation B-n}  

In regions of maser emission, at densities of about $n=10^7cm^{-3}$, 
a field strength of the order of $B=10^3-10^4\mu G$ is observed, 
while in regions of molecular emission, with approximately 
$n=10^2-10^3 cm^{-3}$, the field is found to be of 
the order of $B=10\mu G$ (eg Myers \& Goodman 1988).

A relation of the type $B\propto n^{0.3-0.6}$ may be deduced
from the observations (eg Troland, Crutcher \& Kaz\`{e}s 1986; Heiles 1987; 
Dudorov 1991), 
but it is quite uncertain, especially in the light of the 
above discussion about the intermittency of the distribution of the magnetic 
energy.    
                                                                                                              
The fact remains, however, that observationally the field strength certainly
grows with the gas 
density, and this is found over a range of 6 orders of magnitudes
in density, and 3 orders of magnitude in field strength.

Fig.~4 shows the relation $B-n$ in models A and B.
The dispersion in the relation $B-n$ is shown in the form
of $1-\sigma$ `error' bars in the plot of Fig.~4.
In model A the slope of the relation decreases with time. We find 
$B\propto n^{0.8}$ at  $t=0.2t_{dyn}$, $B\propto n^{0.7}$
at $t=1.0t_{dyn}$, $B\propto n^{0.5}$ at $t=2.0t_{dyn}$, and 
$B\propto n^{0.3}$ at $t=3.0t_{dyn}$. We interpret this time evolution of the
relation $B-n$ in the next section.

In model B there is no significant correlation between $B$ and $n$.
The lack of correlation between $B$ and $n$ was to be expected:
most of the density enhancement occurs by convergent flows 
along the field lines, in the $z$ direction, and these motions do not affect
the magnetic field. A nice illustration of this is provided by
the snapshots in Fig.~5 (lower row), where the density field is stratified
in planes that are predominantly perpendicular
to the $z$ direction (vertical direction in the images). The
densest regions develop at nodes in the large-scale waves, 
as found by Carlberg \& Pudritz (1990).

One could possibly argue that the lack of correlation of $B$ 
and $n$ is consistent 
with the majority of observations, because no field is detected, and 
hence nothing can be said about the correlation.  However, as further
discussed in Section 4 below, the upper limits of the non-detections speak
against this interpretation.  Also, the cases where a field {\em 
is} detected would then have to be explained with {\em ad hoc} arguments,
rather than as a natural part of a statistical distribution.

An interesting consequence of the model A results is that the observed
relation could in principle be used to set an independent estimate of the
age of molecular clouds:
the comparison between model A and the observations would indicate
that molecular clouds, on the average, have survived, since they were
almost uniform in density and field distribution, for 2 or 3 dynamical times,
that is about $10^7$ years, on a scale of $20$ pc. This estimate has little
to do with the estimate of the time-scale for massive stars to destroy the
parent molecular cloud (Blitz \& Shu 1980), since that calculation has 
nothing to say about the lifetime of the cloud after it cooled from 
diffuse warm gas.

To conclude, note that Fig.~4 illustrates that, even if
in model A the mean magnetic field is such that the 
$\langle \beta \rangle\approx 0.7$,
in regions only 10 times denser than the mean, the
field can be as intense as in model B.

\subsection{Cloud and flow topology}

An understanding of the spatial structure of the density field 
in dark clouds
is very important for a correct interpretation of observational data,
and for the formulation of a number of physical models.

The snapshots in Fig.~5 give an idea of the dimensionality
of the structures in the density field. It is apparent that experiment
B (lower row of panels) develops a stratified density field, with sheet-like 
structures, 
the sheets being approximately perpendicular to the magnetic field (the $z$ 
direction
corresponds to the vertical direction in the figures). This is consistent
with the fact that only the motions in the $z$ direction can compress
the gas to very high density.

The topology of the density field in experiment A (upper row of panels
in Fig.~5) has a clear evolution in time. In the very beginning, 
until $t\approx0.4t_{dyn}$, the density
grows predominantly in sheets. These are the fronts of blobs of coherent
motion, advancing at supersonic velocity. Later, these fronts start
to intersect each other, and the density increases especially in 
filaments (at the intersections of fronts). The evolution continues with the
intersection of filaments into knots, at $t\approx1.5t_{dyn}$. The 
fully developed topology, at $t=2.0t_{dyn}$, is characterized by both 
filaments and knots (cores).

The evolution of the magnitudes of the mass density and the magnetic 
field may be discussed with reference to Lagrangian version of the 
continuity equation,
\begin{equation}
\label{Lagrrho}
{{\rm D} \ln \rho \over {\rm D} t} = - \div \vv,
\end{equation}
and the scalar induction equation
\begin{equation}
\label{LagrB}
{{\rm D} \ln |B| \over {\rm D} t} = -\nabla_{\perp}\cdot \vv,
\end{equation}
where $\nabla_{\perp}\cdot$ stands for the divergence perpendicular 
to the magnetic field, following both the motion and the change of 
orientation of the field lines.

Although $-\div\vv$ vanishes for the solenoidal initial condition, the
supersonic motions rapidly lead to the formation of shock fronts, where
the local value of $-\div\vv$ is large and positive because of the 
discontinuity in the velocity perpendicular to the shock.

The initially homogeneous magnetic field is carried along by the 
perpendicular components of the velocity field, and is hence also
collected into sheets, except at those rare locations where the initial
field happens to be strictly parallel to the velocity field.  This explains
why sheets initially form in both mass density and magnetic flux
density, and why the $B-n$ relation initially has an exponent close to unity.

Note that the usual argument for a slope of 2/3, that applies to isotropic
and non-shocking compressible motion does not apply here, because of the
development of discontinuities. In term of Eqs.\ \ref{Lagrrho} and
\ref{LagrB}, the 2/3 follows if $-\nabla_{\perp}\cdot \vv$ typically picks
up two of the three directional contributions to $-\div\vv$. However, at a
shock, the divergence is dominated by the derivative in one particular
direction; the one perpendicular to the shock front. The magnetic field
that is swept into the discontinuity quickly becomes almost parallel to the
shock front, because the component in the plane of the shock grows
exponentially with time. Thus, as long as the topology is dominated by
sheets, the mass density and the magnetic flux grow more or less in unison
in the sheets, corresponding to an exponent in the $B-n$ relation close to
unity.

In the subsequent evolution, there are effects that tend to reduce the
exponent in the $B-n$ relation.   First,  the non-linear evolution of the
initially solenoidal velocity field also leads to the development of 
regions of space with a positive divergence, in which both the mass 
density and the magnetic flux density decline.  In these regions, there
is no particular dominance of the cross-field divergence,  and thus the
three-dimensional divergence picks up un additional contribution relative
to the two-dimensional divergence, consistent with the classical 2/3
argument outlined above.

In experiment B the motions are mainly Alfv\'{e}n waves, and therefore the
velocity is predominantly perpendicular to the direction of the magnetic
field. This is illustrated in Fig.~6, where we have plotted the histogram
of $cos(\alpha)$, where $\alpha$ is the angle between ${\bf v}$ and ${\bf
B}$.   Note that, even though motions across the field lines are the 
most common ones, it is the less frequent motions along the field 
lines that dominate the dissipation, as mentioned in connection with 
the discussion of Fig.~5 above.  The motions across field lines are 
subject to magnetic restoring forces, and do not lead to density
enhancements.  It is the motions along the field lines that lead to 
the sheet like density enhancements visible in Fig.~5. 

In experiment A, the magnetic field is advected by the flow, and the
stretching of field lines instead produces some alignment between ${\bf v}$
and ${\bf B}$, already before one dynamical time has passed, as illustrated in
Fig.~6.

Alignment between $\bb$ and $\vv$ may be caused by two, complementary
effects: 1) {\em Dynamical alignment} is expected when the magnetic energy
approaches and exceeds the kinetic energy; the Lorentz force then forces
the flow to be predominantly along the magnetic field lines. 2) {\em
Kinematic alignment} occurs when a spatially non-uniform velocity field
causes stretching of magnetic field lines, and hence a correlation of $\bb$
and $\vv$. Pure shear, for example, tends to align an embedded magnetic
field with the direction of the flow. 

Motions that are aligned with the magnetic field affect the mass density
without affecting the magnetic flux density. In particular, the non-linear
concentration into first sheets and then filaments due to the interaction
of shock fronts continues into the formation of knots in the density field,
by the convergence of matter flowing along filaments. There is no
corresponding process available to a divergence free vector field such as
the magnetic field; once the field has concentrated into filaments, it
cannot concentrate further; the magnetic field in a filament is insensitive
to flow along the filament.

In the same way that converging flows along the magnetic field may lead to
extreme concentrations of mass, those regions where the flow is diverging
along magnetic field lines may lead to extreme rarefactions of mass,
without affecting the magnetic flux density. In $B-n$ scatter plots, this
corresponds to the development of more extreme excursions of the mass,
relative to those of the magnetic flux density, and hence a flattening of
the $B-n$ relation with time.

In model A, dynamical alignment is at most significant in the few cores that
develop a strong (sub-Alfv\'{e}nic) magnetic field in the early evolution
of the experiment. In scatter plots of $B$ against $n$ most contributions
come from regions where dynamical alignment is unimportant. We thus
conclude that the evolution of the $B-n$ relation in model A, towards a
smaller exponent with time, is caused by the kinematic alignment of $\bb$
and $\vv$.

\subsection{Stellar extinction}

Padoan, Jones, \& Nordlund (1997) have shown that near-infrared
stellar extinction determinations can be used to infer the 3-D
probability distribution of the gas density in dark clouds.
They have shown that there is qualitative and quantitative
agreement between the inferred 3-D density distribution in dark clouds,
and the one produced by their experiment of random supersonic
flows. 

The method is based on the plot of the dispersion of the extinction measurements
in cells, versus the mean extinction in the same cells (Lada et al.\ 1994). 
In Padoan, Jones \& Nordlund (1997), the theoretical 
plots are produced by generating random density fields of given statistics and 
power
spectra. Here we show one example of the same plot, but produced
directly from the density field of experiments A and B. The plot is shown in
Fig.~7, that can be compared directly with the plots shown in Padoan,
Jones, \& Nordlund. The plot of model A looks like the observational one.  
It has a smaller slope because the rms Mach number in Experiment A is smaller
than in the observed cloud.

The plot for Experiment B seems instead to differ from the observational one. 
The absence of locations with very low extinction in model B is due to the 
fact that the density field in experiment B is mainly organized in sheets, 
perpendicular to the magnetic field direction.

\section{Discussion}

It is difficult to get an objective view on the magnetic 
field strength in dark clouds from the literature. The reasons are the 
following:

\begin{itemize}

\item Negative results from observational programs, in which detections 
have been reported, often remain unpublished.

\item The positions searched for Zeeman splitting never represent a
statistically meaningful sample. Favorable regions are always selected,
because the observations are very time consuming.

\item The total number of regions in dark clouds, for which OH Zeeman 
observations are published, is still small.

\end{itemize} 

The best study of OH Zeeman splitting in dark clouds we are aware of is
by Crutcher et al.\ (1993). They selected 4 cores in the Taurus dark cloud 
complex, 2 in the Libra complex, 2 in $\rho$ Oph, 1 in the Orion molecular
ridge, 1 position in L889, and the core of B1 (Barnard 1), in the Perseus region.
The only certain detection is in
the cloud B1. For the other regions, the weighted average value of the field is
$+2.7\pm 1.5 \mu G$ in Taurus, $-2.1\pm 2.8 \mu G$ in Libra, $+6.8\pm 2.5\mu G$
in $\rho$Oph, $-0.6\pm2.1\mu G$ in L889, and $-4.7\pm3.5 \mu G$ in L1647.

Two well known regions of very intense magnetic field are Orion A and 
Orion B. OH Zeeman splitting in Orion A revealed a magnetic field strength
of $B=-125\mu G$ (Troland, Crutcher, \& Kaz\`{e}s 1986), and 
$B=+38\mu G$ toward Orion B (Crutcher, \& Kaz\`{e}s 1983).

Crutcher et al.\ (1996) have reported the first attempt to measure CN 
Zeeman splitting. They observed towards the cores OMC-N4 and S106-CN.
They did not detect the magnetic field, and their upper limits are several
times smaller than the value expected if the observed motions were 
sub-Alfv\'{e}nic.

An average field of $+9\mu G$ was found in Cas A by Heiles and Stevens (1986).

The OH Zeeman splitting should probe regions with $n=10^3-10^4$ cm$^{-3}$,
while the CN Zeeman splitting, should probe regions of $n\approx10^6$ cm$^{-3}$.

Note that all works are biased toward regions of strong field, because only 
regions considered to be favorable for magnetic field detections are selected
for Zeeman splitting observations. Despite that, in many cases the field is not
detected, and upper limits are quite stringent. 

In our model A we find that, if the mean field strength is 
$\langle B^2\rangle^{1/2}=5\mu G$, and the mean density $\langle n\rangle=10^3 
cm^{-3}$, at the 
time $t=2.0t_{dyn}$, then $20\%$ of the total mass is in dense clumps, 
with density ten times larger than the mean, $n\approx10^4 cm^{-3}$, and 
field strength five times, or more, larger than the mean, $B\ge25\mu G$. 
Therefore, even if field strengths of about $30\mu G$ are detected 
sometimes in dense cores, the mean Alfv\'{e}n velocity in the molecular cloud
may be just comparable to the sound speed, $\langle v_{A}^2\rangle^{1/2}\approx 
C_{S}$.

On the basis of the observational results by Crutcher et al.\ (1993, 1996), 
it may be concluded that model A is in consistent with the observational
estimates of magnetic field strength in dark clouds.  The particular values
of the magnetic field used in the comparison here should not be taken too
literally; the small scale field strengths could be larger than estimated by the
Zeeman effect if the magnetic field is tangled. 

It is difficult to envisage how the measurements could be consistent with
model B, however, since the magnetic field in a sub-Alfv\'{e}nic model by
definition has an energy that exceeds the kinetic energy. As demonstrated by
Galsgaard \& Nordlund (1996), a magnetically dominated plasma is able to
quickly dissipate structural complexity, independent of the value of the
resistivity. Thus, a sub-Alfv\'{e}nic field could not remain strongly
tangled, and hence could not avoid detection. Using the same argument, we
expect those cores where a strong (sub-Alfv\'{e}nic) field has indeed been
observed to have a relatively simple magnetic field structure.

Model A can be used for the description of a typical molecular cloud with
linear size of about $5$ pc and $\langle n\rangle=500$ cm$^{-3}$. At the time
$t=1.0t_{dyn}$, the relation $B-n$ is then:

\begin{equation}
B=5\mu G(\frac{n}{500 cm^{-3}})^{0.7}
\end{equation}
Since the exponent is $>0.5$, most of the dense cores 
are found to have magnetic pressure larger than thermal pressure, at early times.
This is illustrated in Fig.~8, where we have plotted the Alfv\'{e}n velocity 
versus
the gas density, after one dynamical time, for experiments A and B. In model A,
one can see that, on average, regions with density larger than the mean are 
characterized
by an Alfv\'{e}n velocity larger than the sound velocity, and therefore by a 
magnetic 
pressure larger than the thermal pressure.

The lowering of the exponent in the $B-n$ power law with time means that 
later on, the dominance of the magnetic field in these cores tends to be
reduced. Although the flow is random and approximately isotropic, the 
kinematic alignment of ${\bf B}$ with ${\bf v}$ makes dense cores accrete 
mass along ${\bf B}$ at an increased rate. Therefore, the accretion of mass
around dense cores, embedded in a random flow with $\beta\approx 1$, is such that,
while magnetic pressure becomes dominant over thermal pressure during the initial
phase of fragmentation, later on magnetic pressure decreases to the level of
thermal pressure, even in the absence of gravity, 
and on a timescale that is competitive with ambipolar diffusion. This 
mechanism could be relevant for the process of star formation.

The main point we want to clarify in this paper is that the fact of finding
some cloud cores, with sub-Alfv\'{e}nic velocity dispersions and with magnetic 
pressure larger than thermal pressure, does not
necessarily mean that the dynamics of molecular clouds is dominated by
MHD waves; those cores may be formed, in a few million years, in a 
supersonic and super-Alfv\'{e}nic flow, only marginally affected by the
magnetic field (model A). The dynamics becomes strongly affected by the magnetic
field only in some very dense regions, on small scales, and at early times.

We stress that a key theoretical ingredient to the interpretation
of OH Zeeman splitting data is the $B-n$ relation. The fact that the 
exponent of the relation is $>0.5$ for more than two dynamical times
is the reason why sub-Alfv\'{e}nic cores can be found in the experiment.

\section{Conclusions}

In this work we have shown that:

\begin{itemize}

\item Supersonic motions in model A are relatively long-lived, with decay
times comparable to the 
estimated life-time of molecular clouds. Even in the absence of an energy
source, supersonic and super-Alfv\'{e}nic motions can persist as such, on a 
scale of $20$ pc, for about $10^7$ years.

\item Supersonic sub-Alfv\'{e}nic motions (model B) dissipate their energy 
almost as fast as super-Alfv\'{e}nic motions.
The original motivation for the MHD wave model is therefore absent.

\item Random supersonic motions produce a very intermittent probability 
distribution of the magnetic energy, with an exponential tail. Since the 
distribution
is so intermittent, one could easily detect a field strength that
is several times in excess of the mean strength. 

\item Dense cores, with magnetic pressure larger than thermal pressure,
and velocity dispersions smaller than $v_{A}$, are found as the result of 
the evolution of supersonic and super-Alfv\'{e}nic flows (model A). 

\item A power law statistical $B-n$ relation 
is generated by supersonic motions, in model A, but not by
MHD waves, in model B. The exponent of the relation is $>0.5$ for about two
dynamical times, which allows for the existence of cores with $v_{A}>C_{S}$.

\item The exponent in the $B-n$ relation decreases with time, because  
magnetic field lines are stretched and partially aligned with the flow.  
The statistical importance of the magnetic pressure in dense cores is thus 
expected to decrease with time, even in the absence of gravity and ambipolar 
diffusion.

\item The topology of the density field generated by model A
is mainly filamentary and clumpy. This is reminiscent of the
morphology of molecular clouds. On the other hand, the density
field produced by MHD waves in model B is mainly structured in sheets, 
perpendicular to the magnetic field.

\item The statistical properties of the density field generated by random 
supersonic motions (as in model A) are in agreement with the statistical 
properties 
of the gas density in dark clouds, as inferred from stellar extinction 
determinations. Model B does not give such a good agreement.

\end{itemize}

The comparison between model A and model B shows that model A
(super-Alfv\'{e}nic motions) provides a nice understanding of
the dynamics and structure of molecular clouds. Moreover, it is
consistent with the OH Zeeman detections.

We conclude that the supersonic motions observed in dark clouds,
on a scale of a few parsecs to several parsecs, 
can be understood, without any major problem, as super-Alfv\'{e}nic 
motions, although sub-Alfv\'{e}nic dense cores certainly exist.

\acknowledgements

We thank Prof. L. Mestel and Prof. E. Zweibel for their comments on the 
manuscript.
This work has been supported by the Danish National Research Foundation
through its establishment of the Theoretical Astrophysics Center.
Computing resources were provided by the Danish National Science 
Research Council, and by the French `Centre National de Calcul 
Parall\`{e}le en Science de la Terre'.

\clearpage
\begin{figure}
\begin{center}
  \begin{minipage}{80mm}
    \centerline{
    \epsfxsize=80mm
    \epsfbox{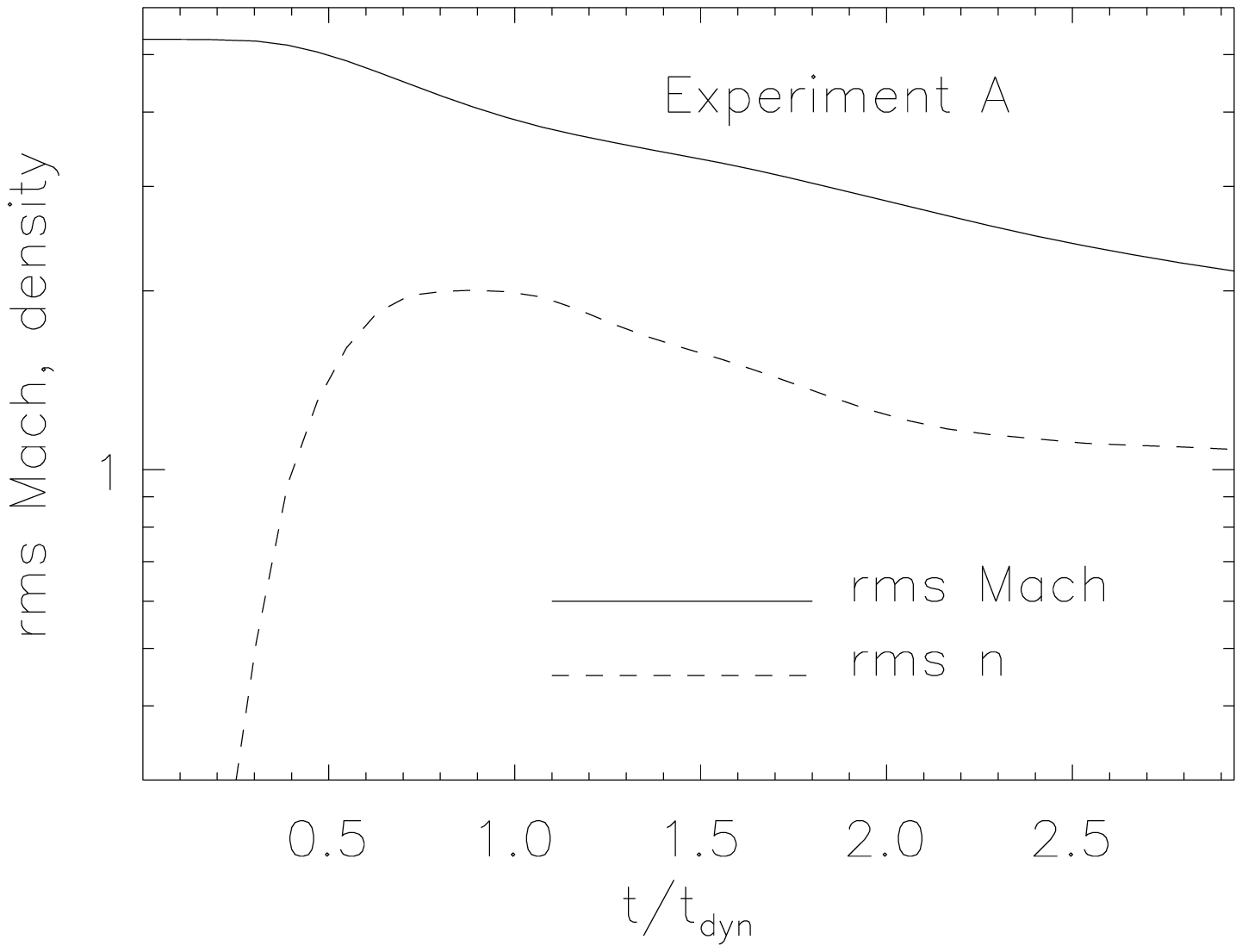}
    }
  \end{minipage}
  \begin{minipage}{80mm}
    \centerline{
    \epsfxsize=80mm
    \epsfbox{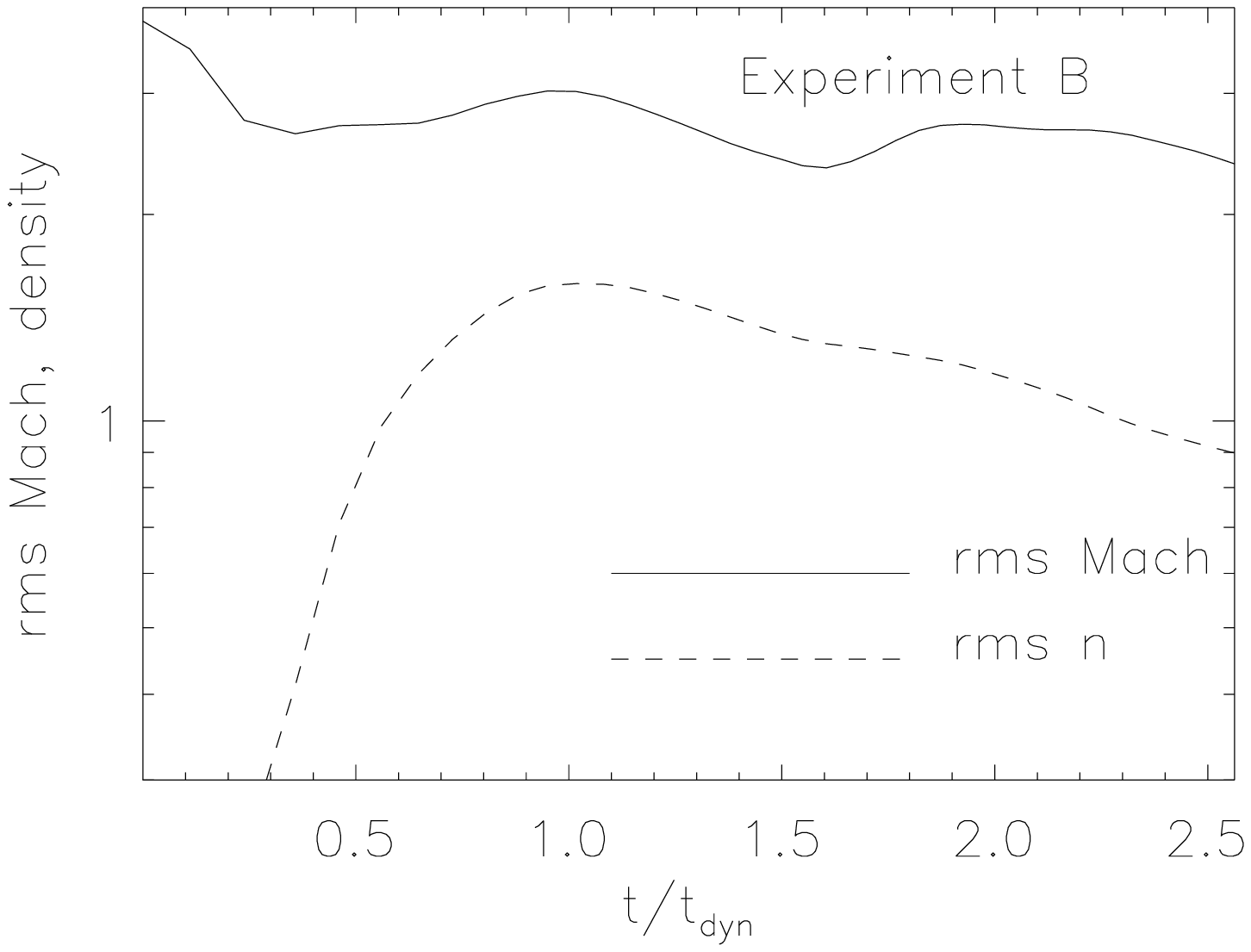}
    }
  \end{minipage}
\caption{Time evolution of rms Mach number and rms number density.}
\end{center}
\end{figure}

\clearpage
\begin{figure}
\label{fig2}
\begin{center}
  \begin{minipage}{80mm}
    \centerline{
    \epsfxsize=80mm
    \epsfbox{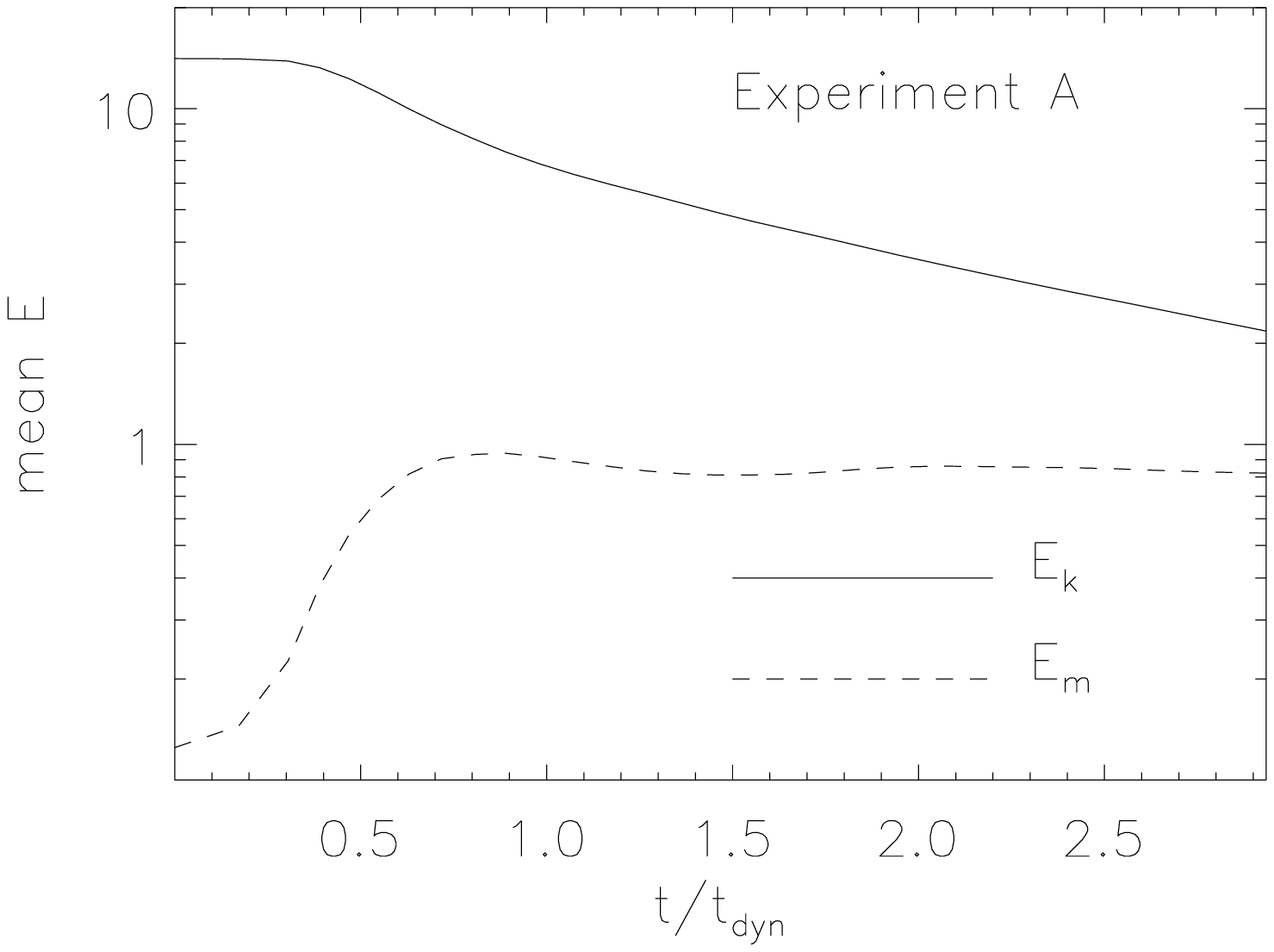}
    }
  \end{minipage}
  \begin{minipage}{80mm}
    \centerline{
    \epsfxsize=80mm
    \epsfbox{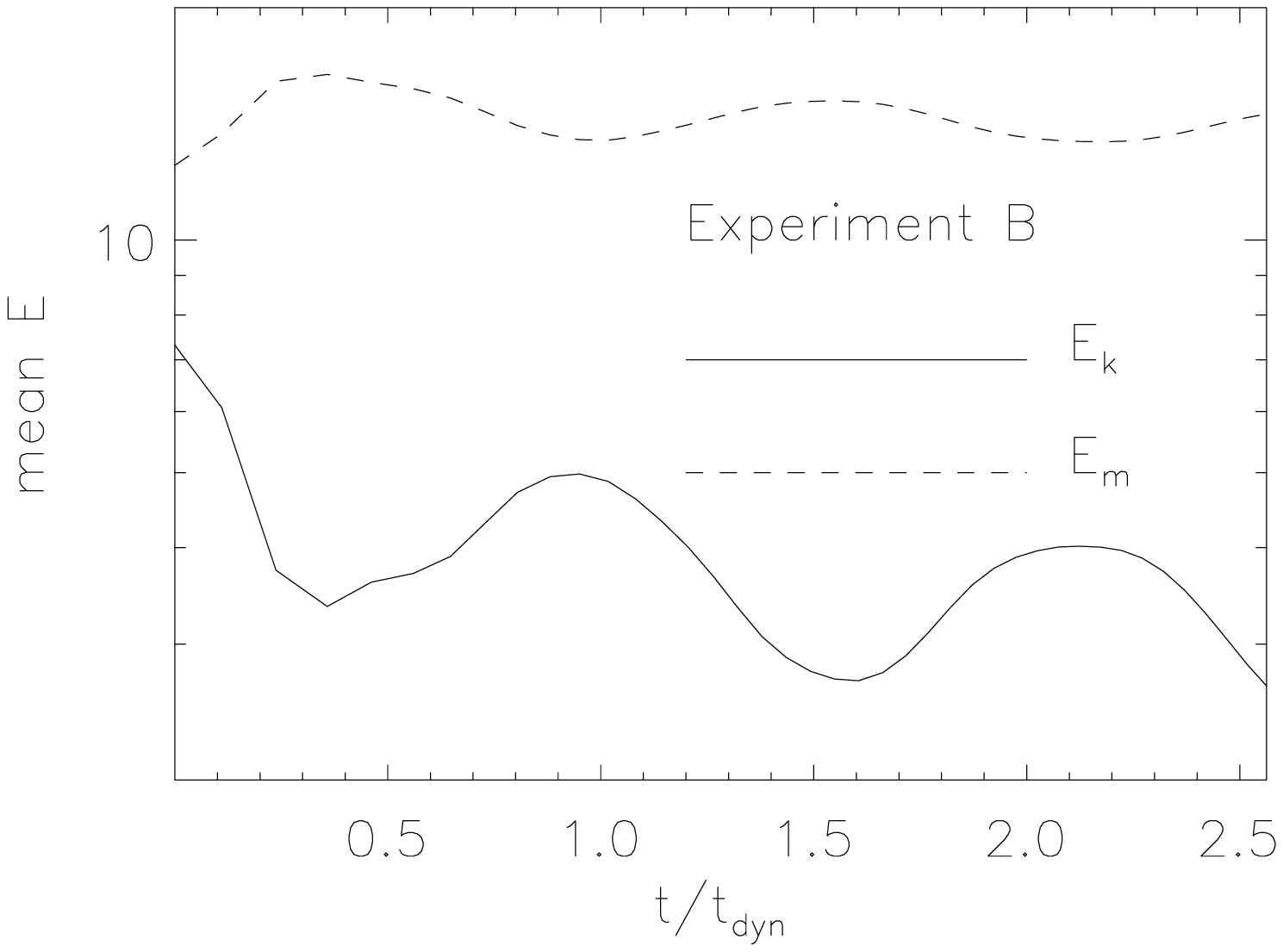}
    }
  \end{minipage}
\caption{Time evolution of kinetic and magnetic energies, in units of the
mean thermal energy.}
\end{center}
\end{figure}

\clearpage
\begin{figure}
\begin{center}
  \begin{minipage}{80mm}
    \centerline{
    \epsfxsize=80mm
    \epsfbox{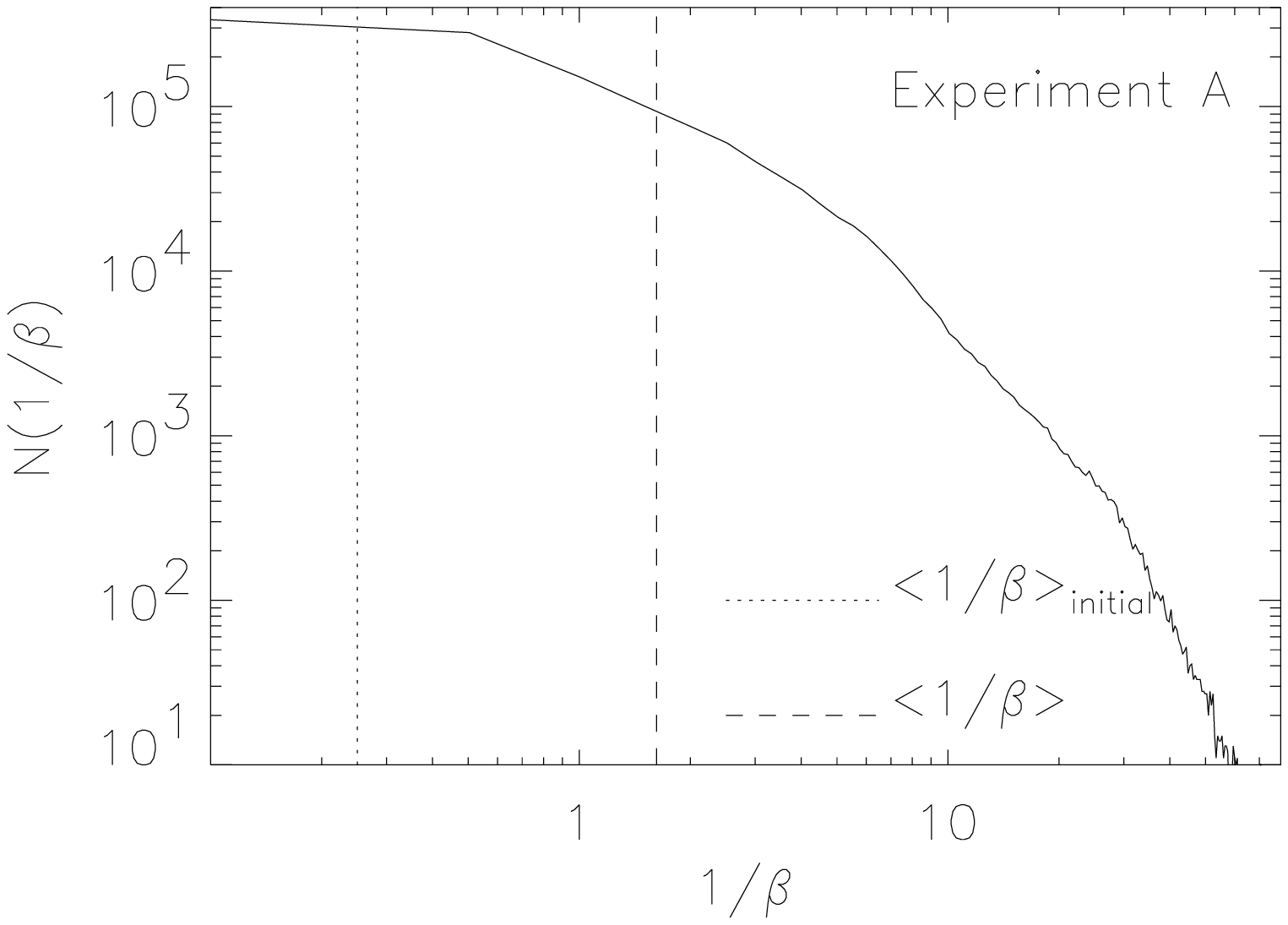}
    }
  \end{minipage}
  \begin{minipage}{80mm}
    \centerline{
    \epsfxsize=80mm
    \epsfbox{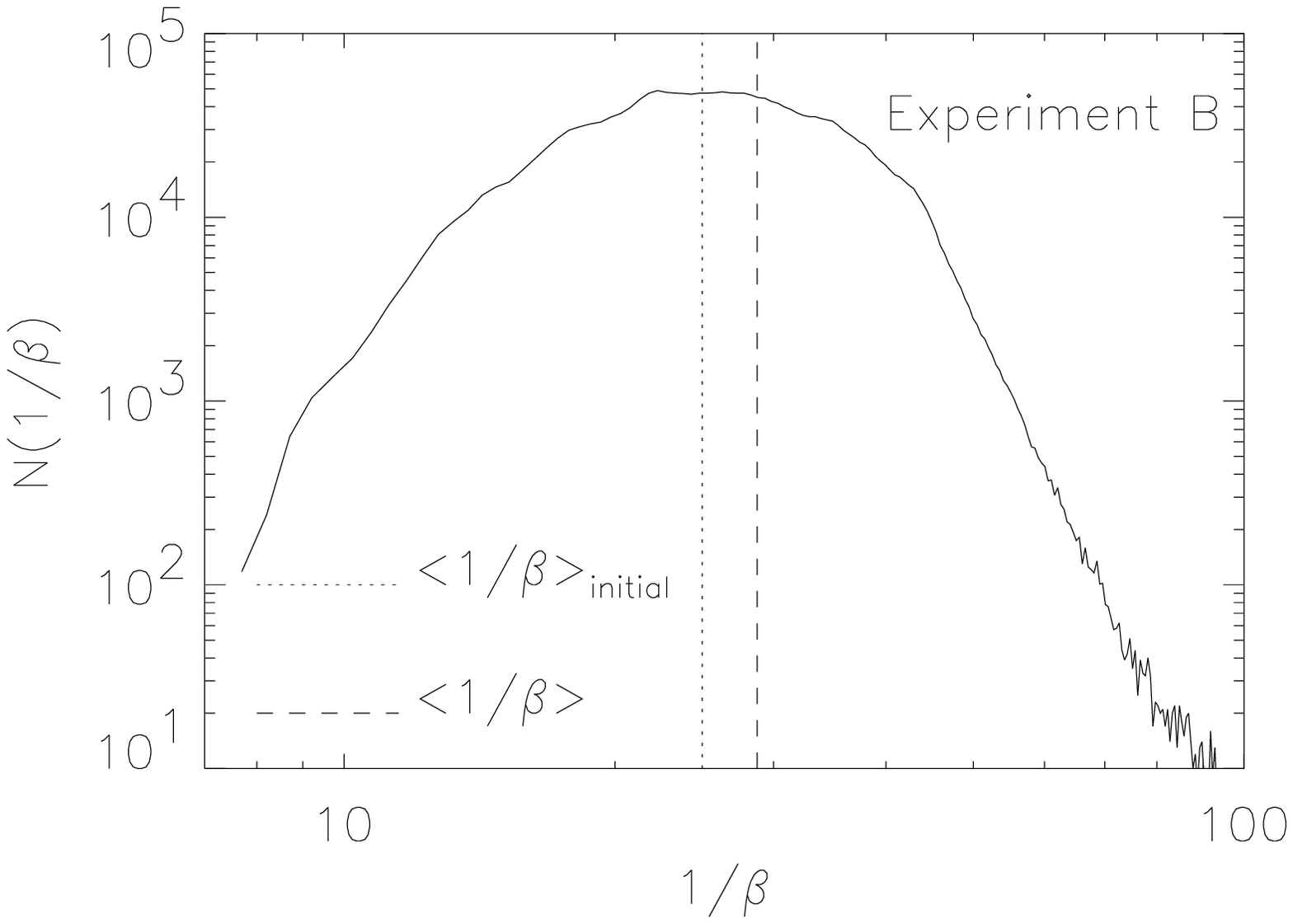}
    }
  \end{minipage}
\caption[]{Distribution of magnetic energy in units of the mean gas pressure,
after one dynamical time. The dotted vertical line marks the initial mean value of
the magnetic energy, and the dashed line the mean value after one dynamical time.}
\end{center}
\end{figure}

\clearpage
\begin{figure}
\begin{center}
  \begin{minipage}{80mm}
    \centerline{
    \epsfxsize=80mm
    \epsfbox{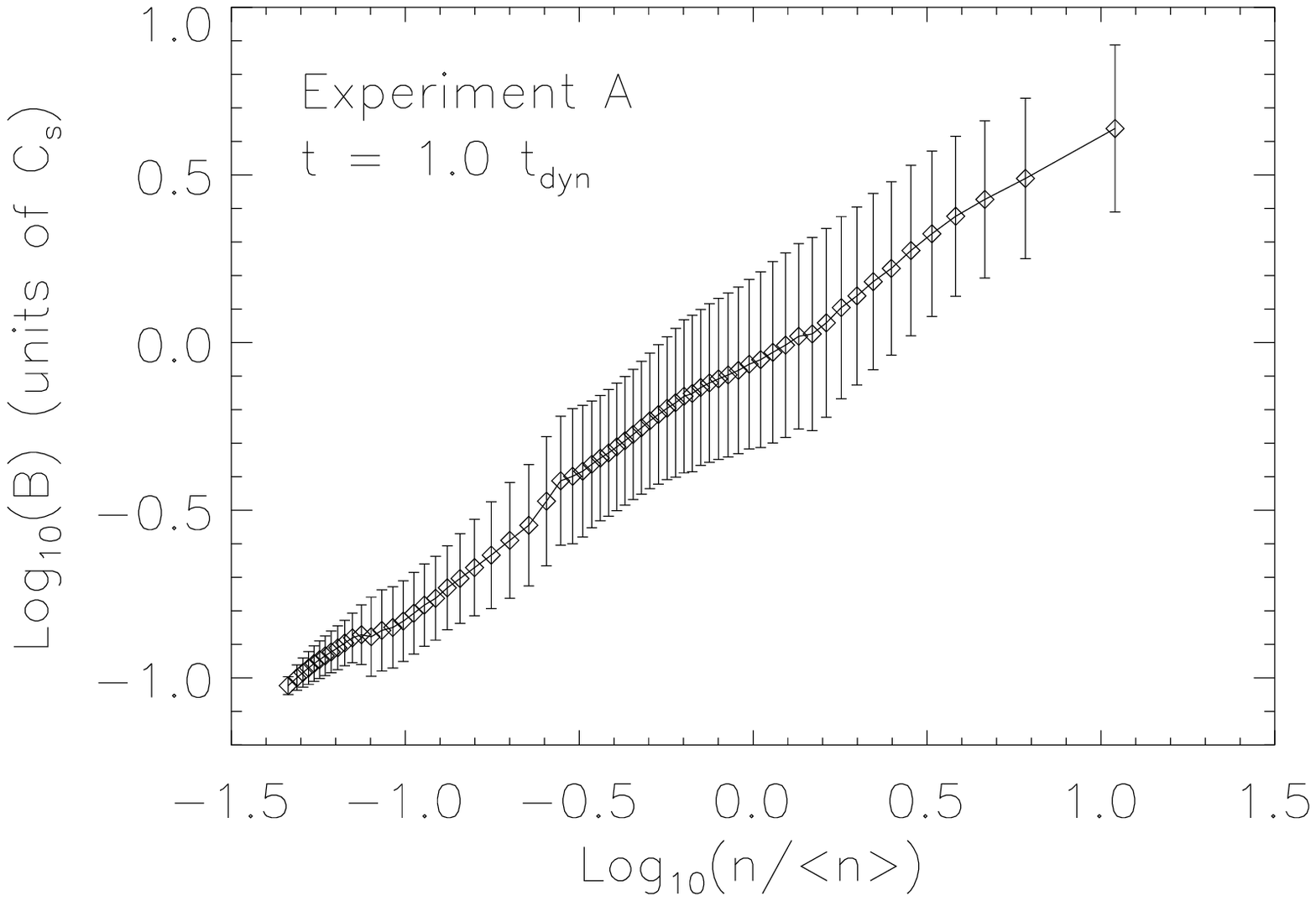}
    }
  \end{minipage}
  \begin{minipage}{80mm}
    \centerline{
    \epsfxsize=80mm
    \epsfbox{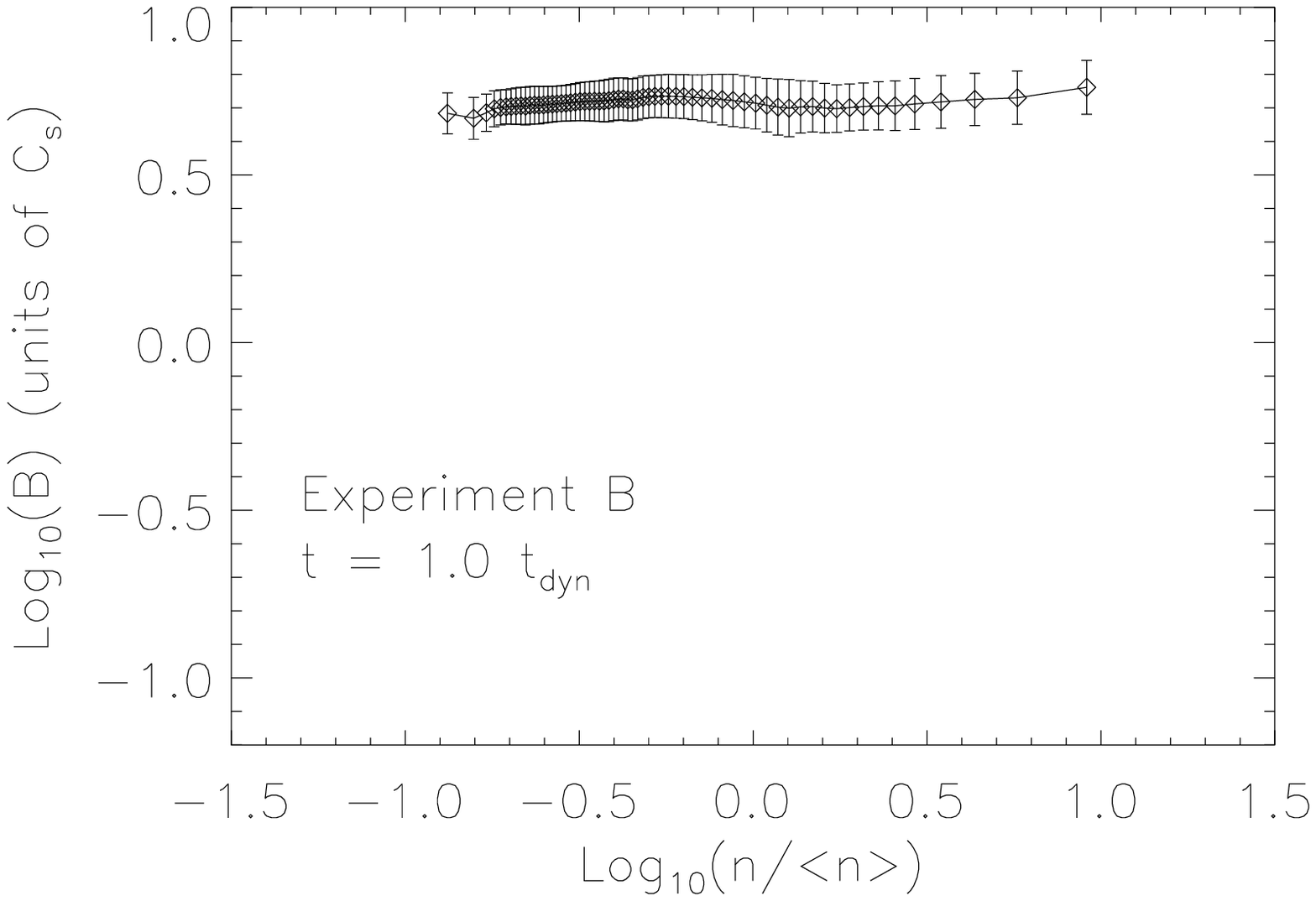}
    }
  \end{minipage}
\caption[]{The relation $B-n$, after one dynamical time. The magnetic field is
expressed in the numerical units, that is as an Alfv\'{e}n velocity (defined with 
the 
mean gas density), in units of the sound speed. The $1-\sigma$ 
`error' bars show the dispersion of values of $B$ around the mean for each 
bin. The size of the bins is chosen in order to have the same number of 
measurements inside each bin.}
\end{center}
\end{figure}

\clearpage
\begin{figure}
\label{fig3}
\begin{center}
  \begin{minipage}{50mm}
    \centerline{
    \epsfxsize=50mm
    \epsfbox{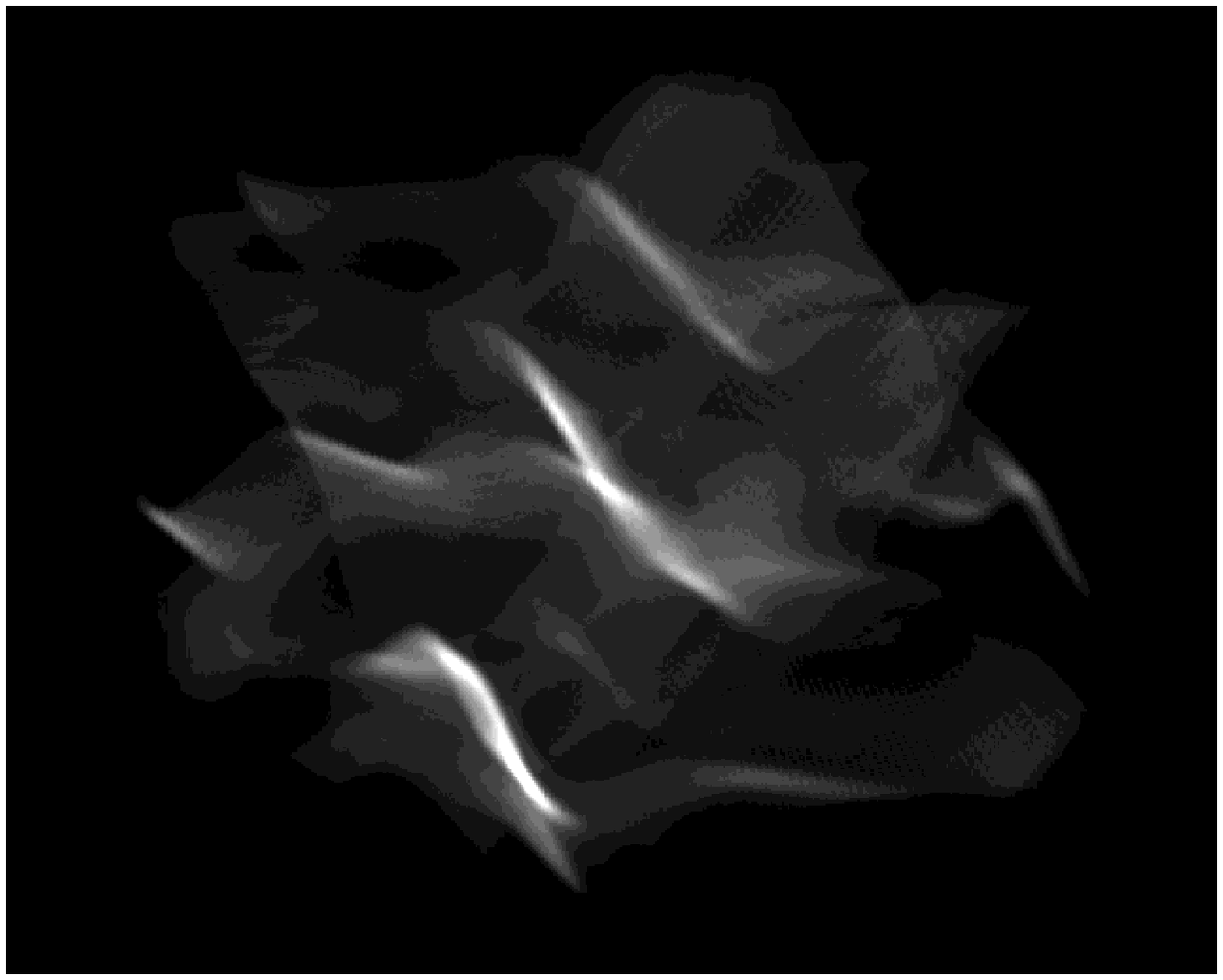}
    }
  \end{minipage}
  \begin{minipage}{50mm}
    \centerline{
    \epsfxsize=50mm
    \epsfbox{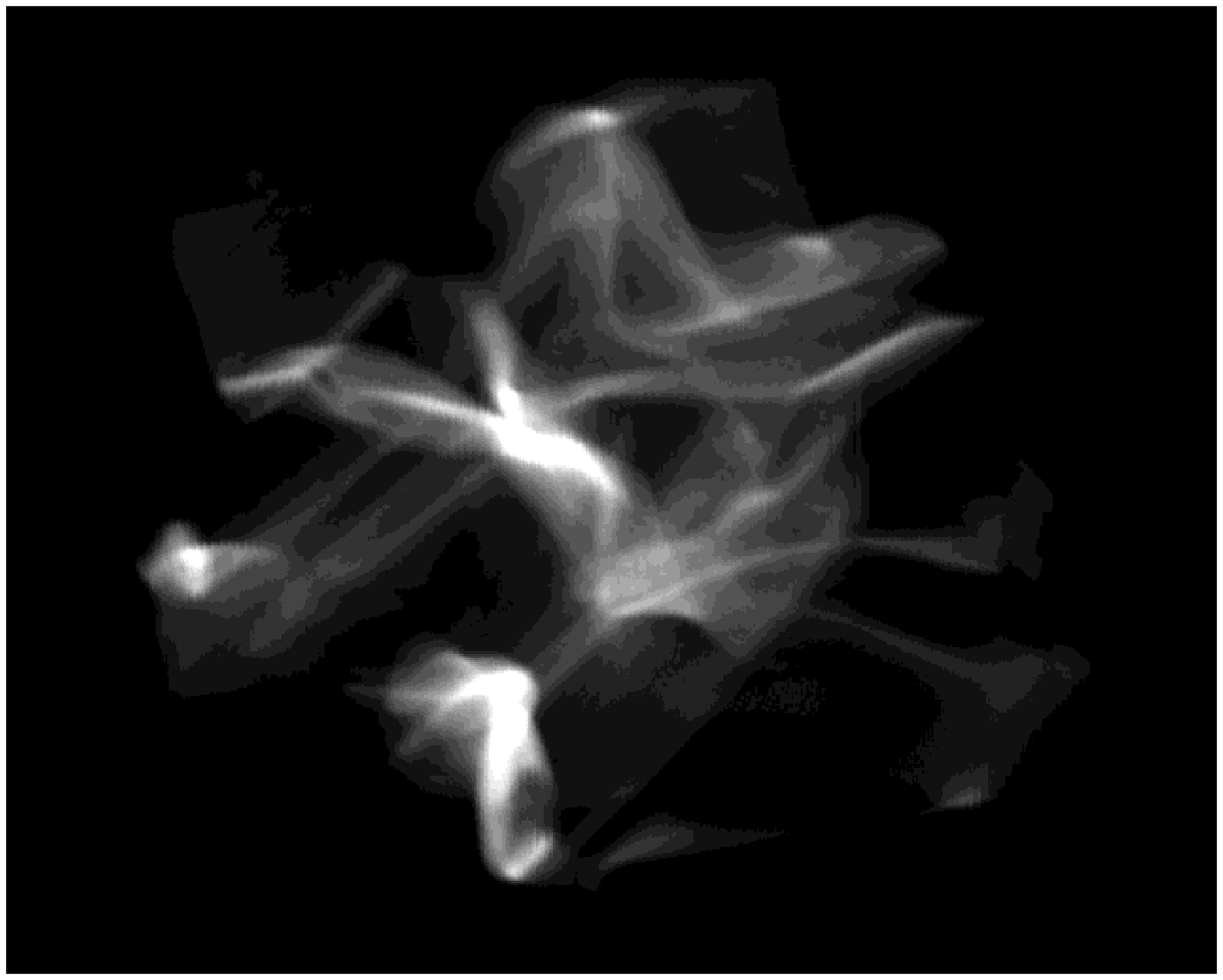}
    }
  \end{minipage}
  \begin{minipage}{50mm}
    \centerline{
    \epsfxsize=50mm
    \epsfbox{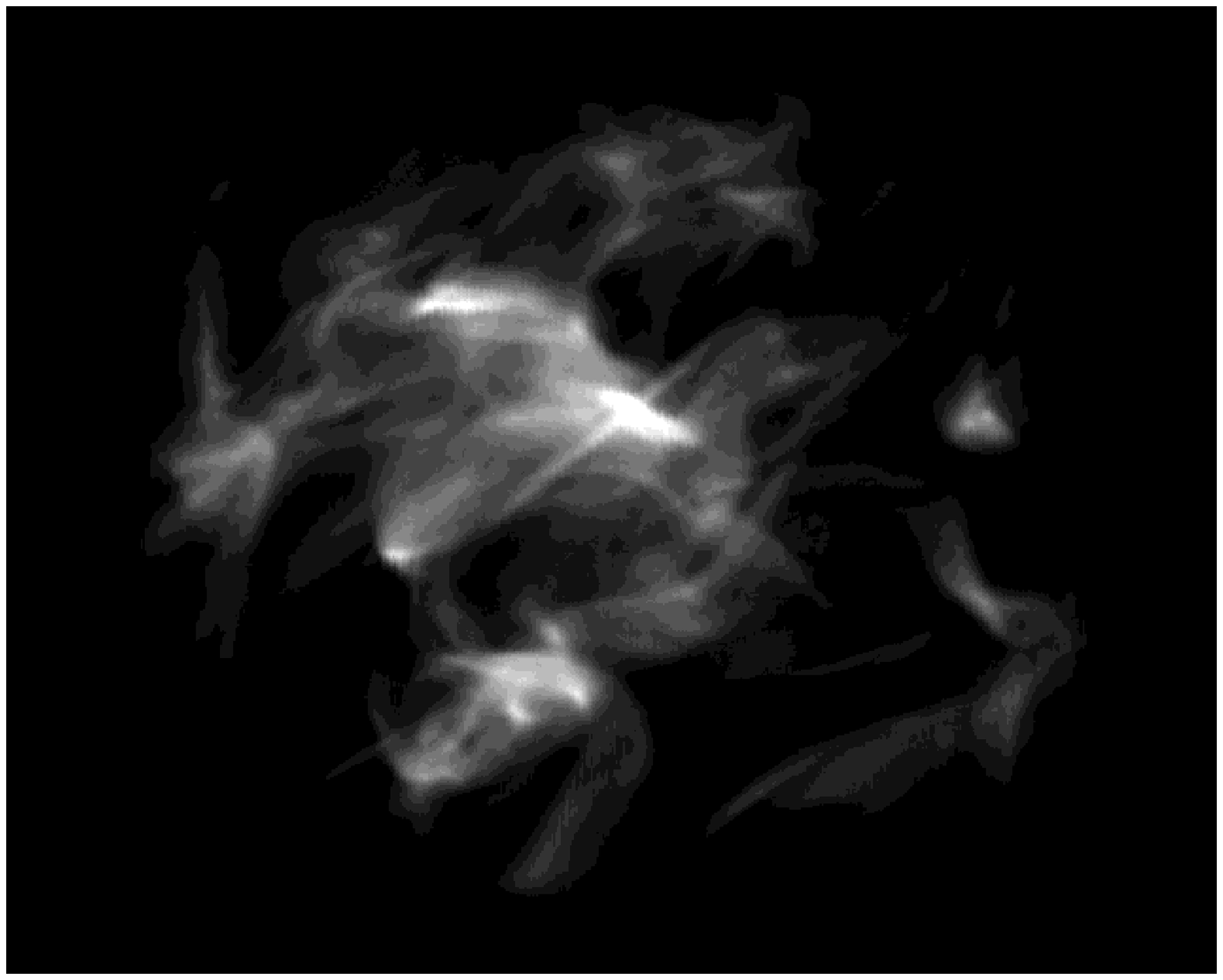}
    }
  \end{minipage}
\begin{minipage}{50mm}
    \centerline{
    \epsfxsize=50mm
    \epsfbox{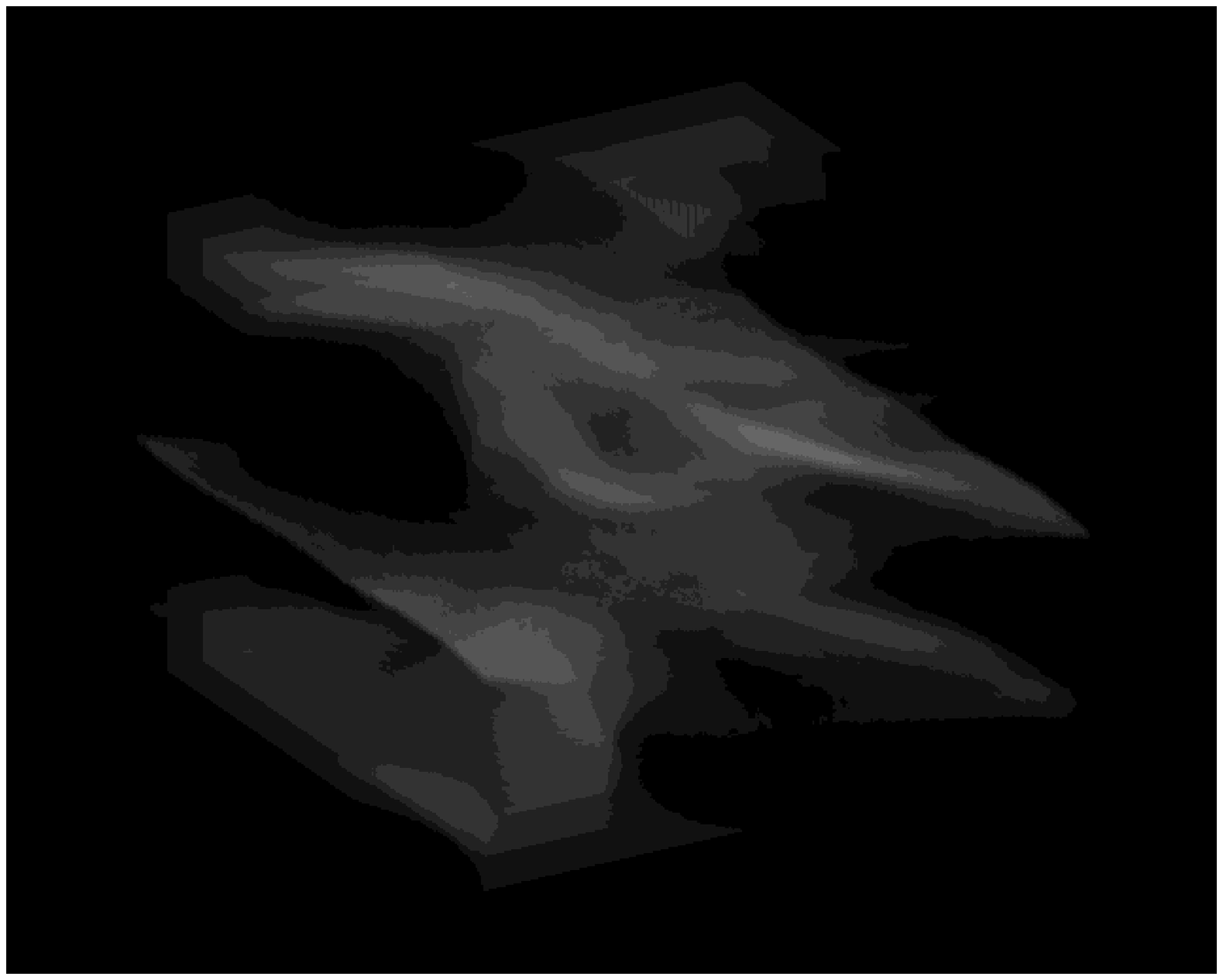}
    }
  \end{minipage}
  \begin{minipage}{50mm}
    \centerline{
    \epsfxsize=50mm
    \epsfbox{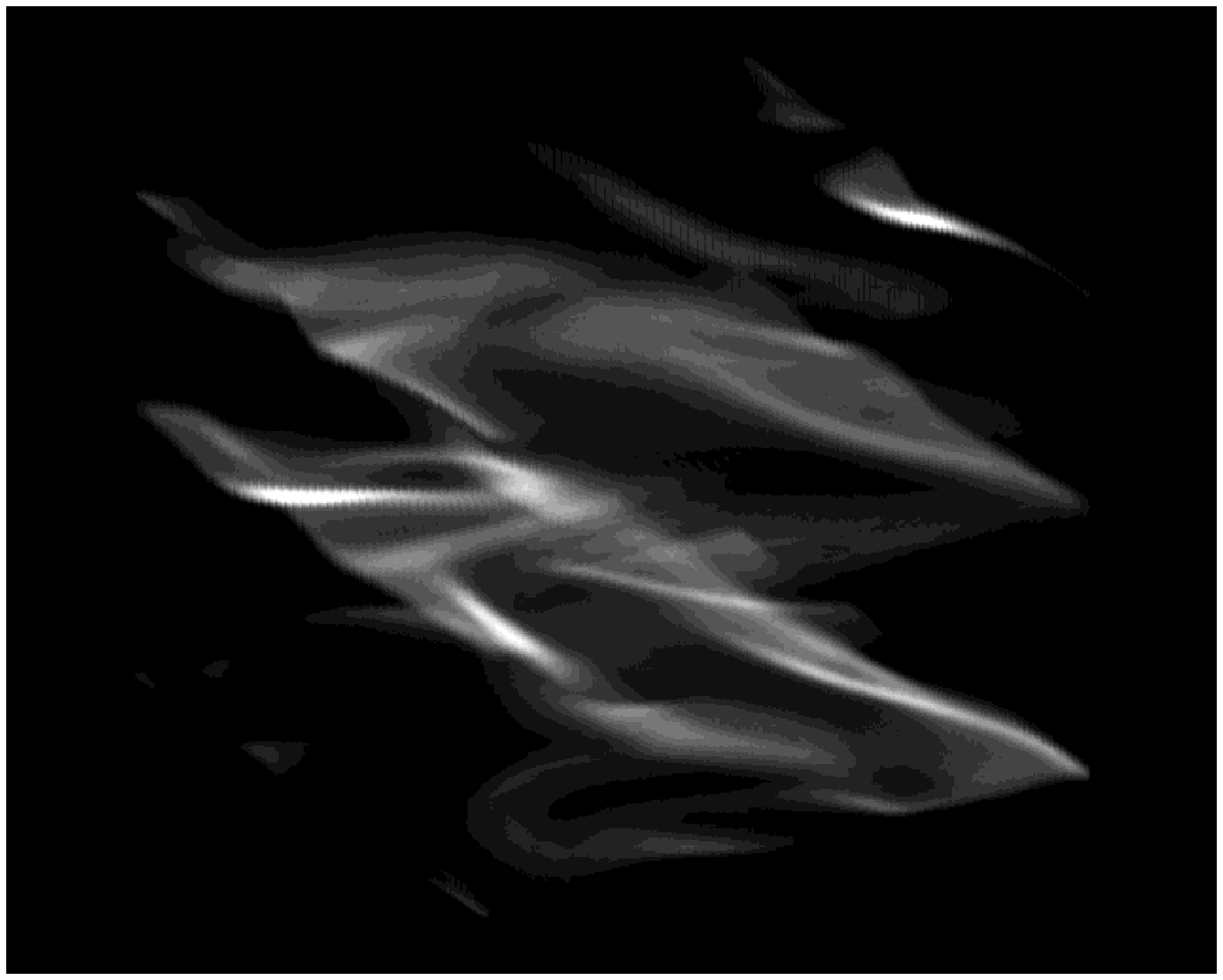}
    }
  \end{minipage}
  \begin{minipage}{50mm}
    \centerline{
    \epsfxsize=50mm
    \epsfbox{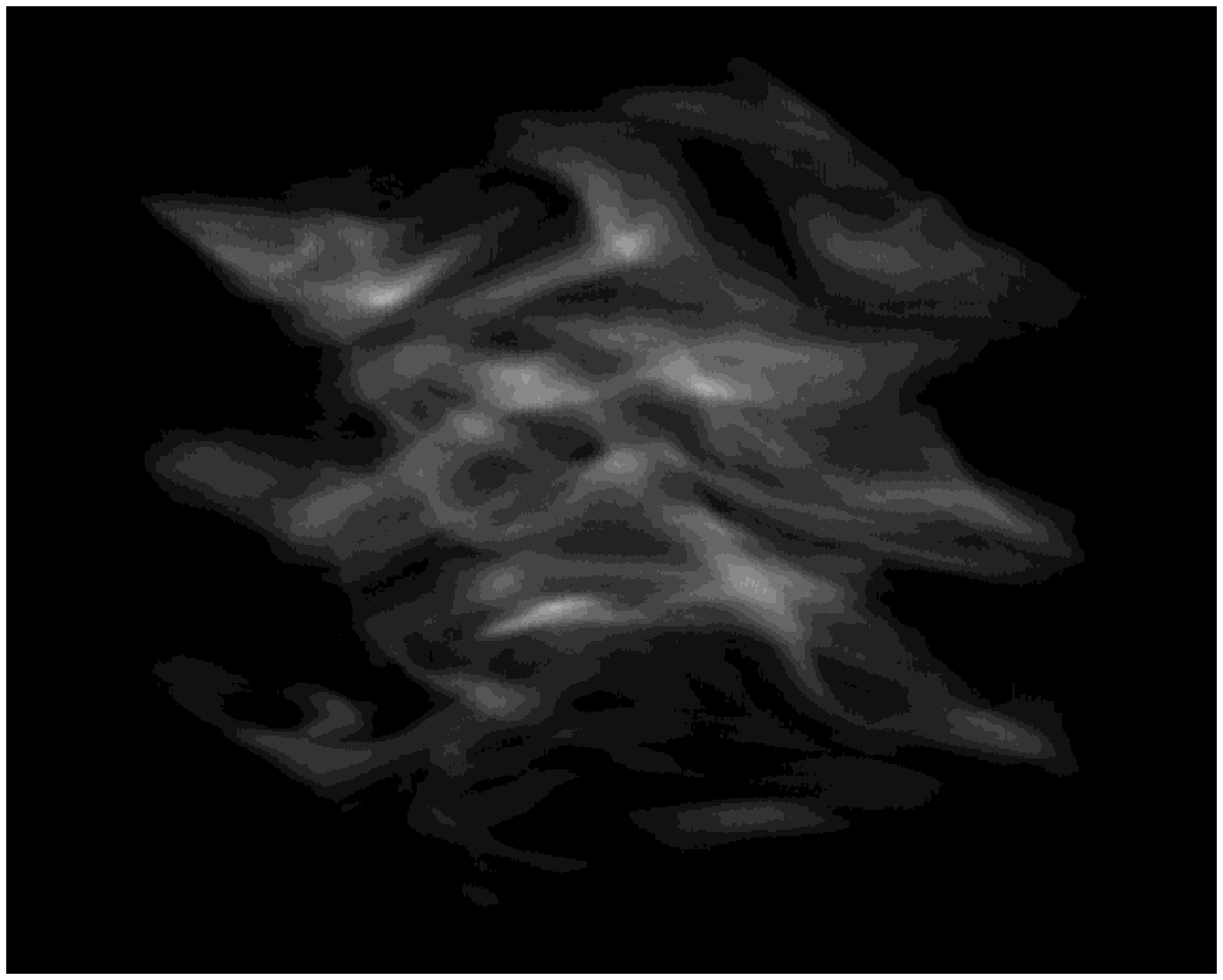}
    }
  \end{minipage}
\caption[]{3-D representation of the time evolution of the density field.
Time evolves from left to right.  The upper row of panels show model A, 
while the lower row shows model B.}
\end{center}
\end{figure}

\clearpage
\begin{figure}
\begin{center}
  \begin{minipage}{80mm}
    \centerline{
    \epsfxsize=80mm
    \epsfbox{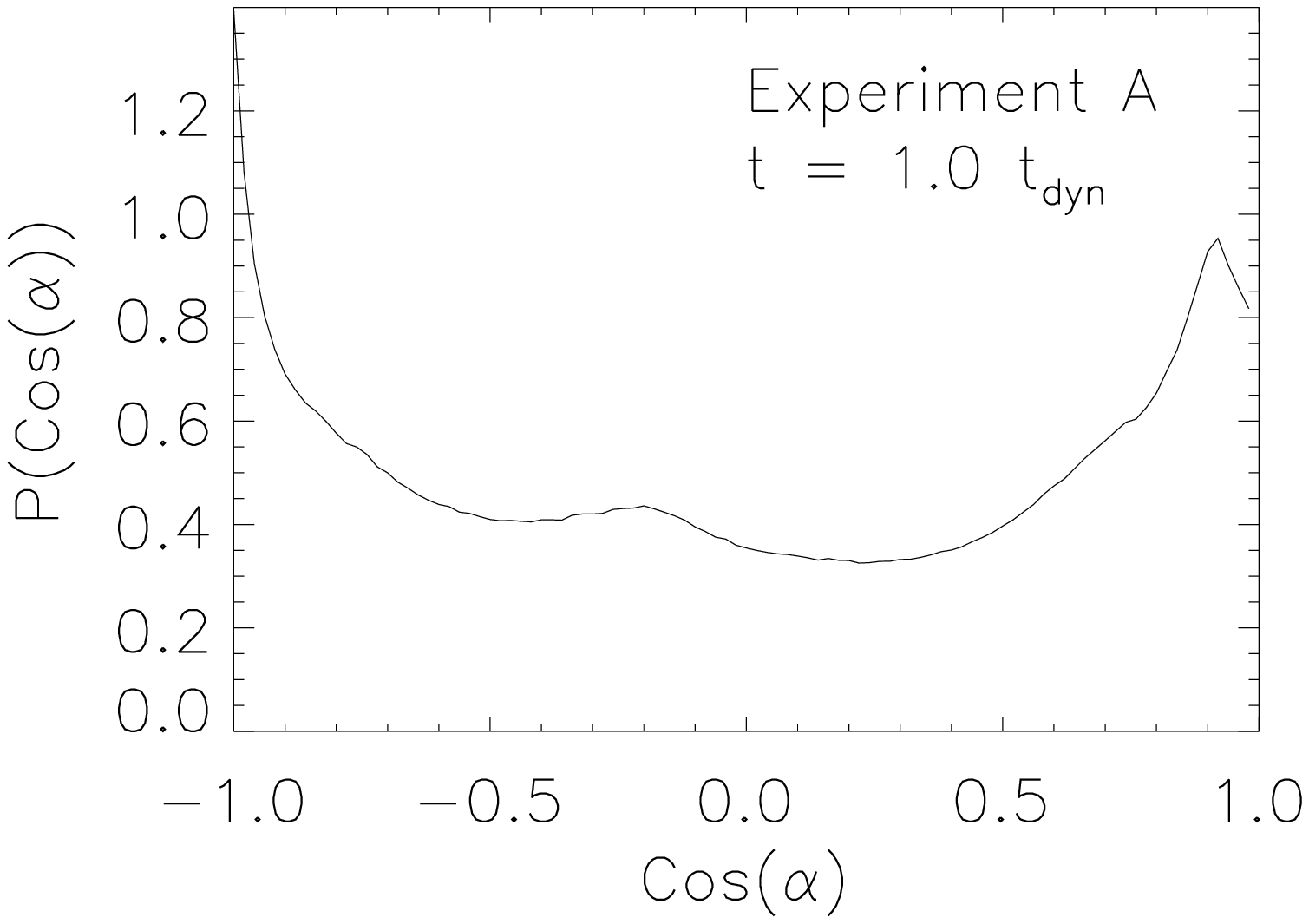}
    }
  \end{minipage}
  \begin{minipage}{80mm}
    \centerline{
    \epsfxsize=80mm
    \epsfbox{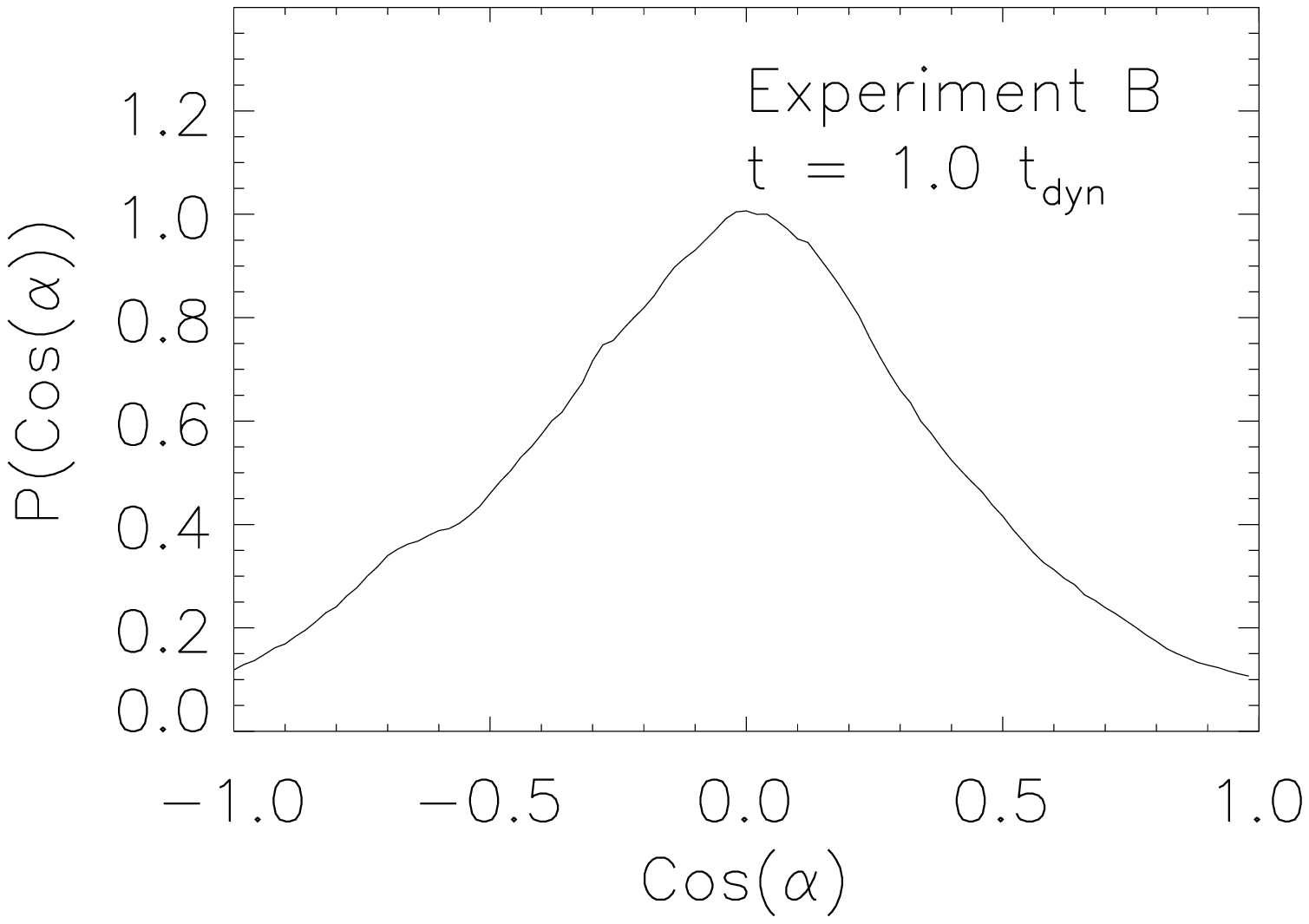}
    }
  \end{minipage}
\caption[]{Histograms of the cosine of the angle between ${\bf v}$ and ${\bf B}$. 
In 
experiment A there is a partial alignement, while in experiment B the two fields 
are 
mainly perpendicular to each other.}
\end{center}
\end{figure}

\clearpage
\begin{figure}
\begin{center}
  \begin{minipage}{80mm}
    \centerline{
    \epsfxsize=80mm
    \epsfbox{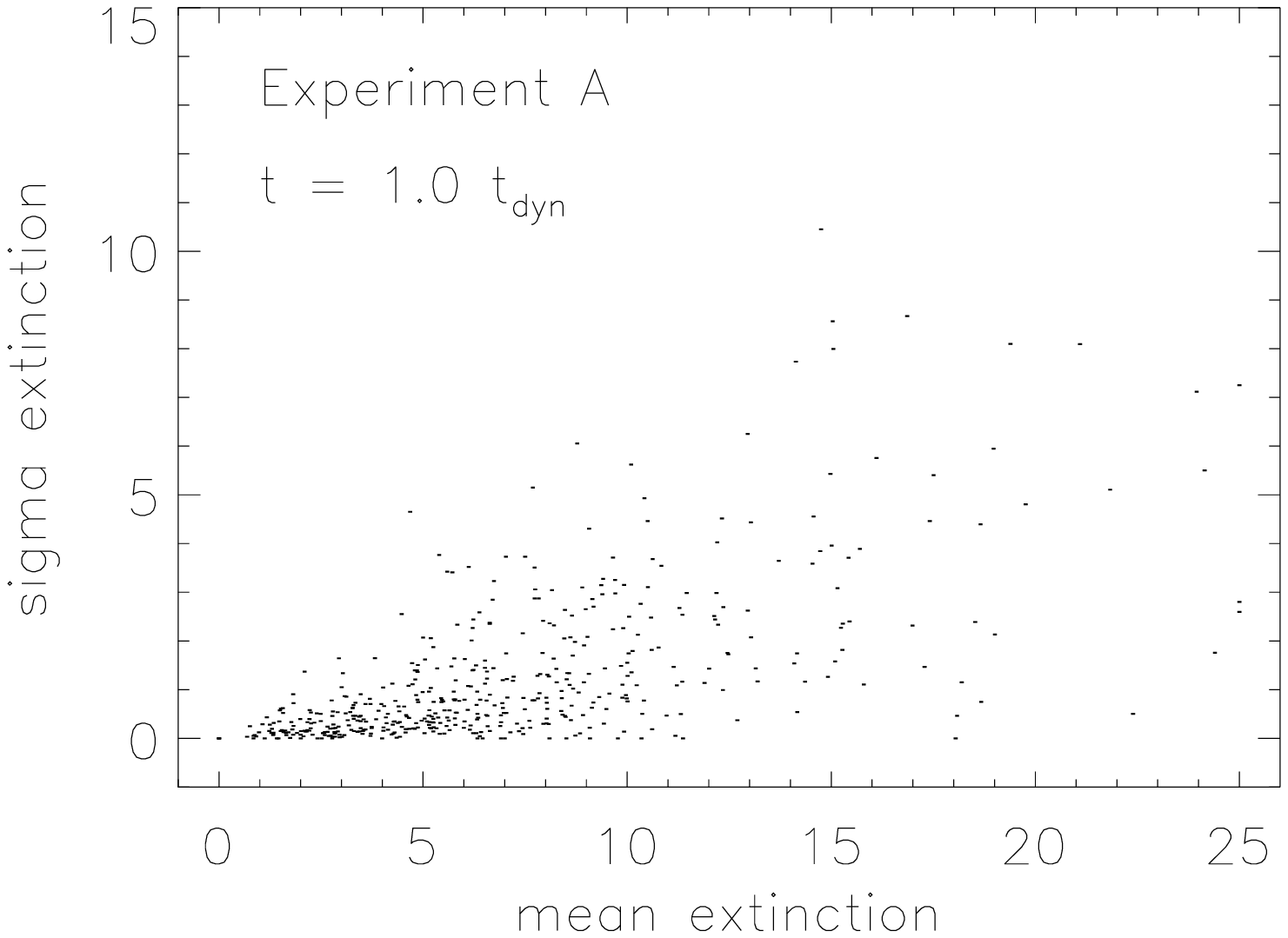}
    }
  \end{minipage}
  \begin{minipage}{80mm}
    \centerline{
    \epsfxsize=80mm
    \epsfbox{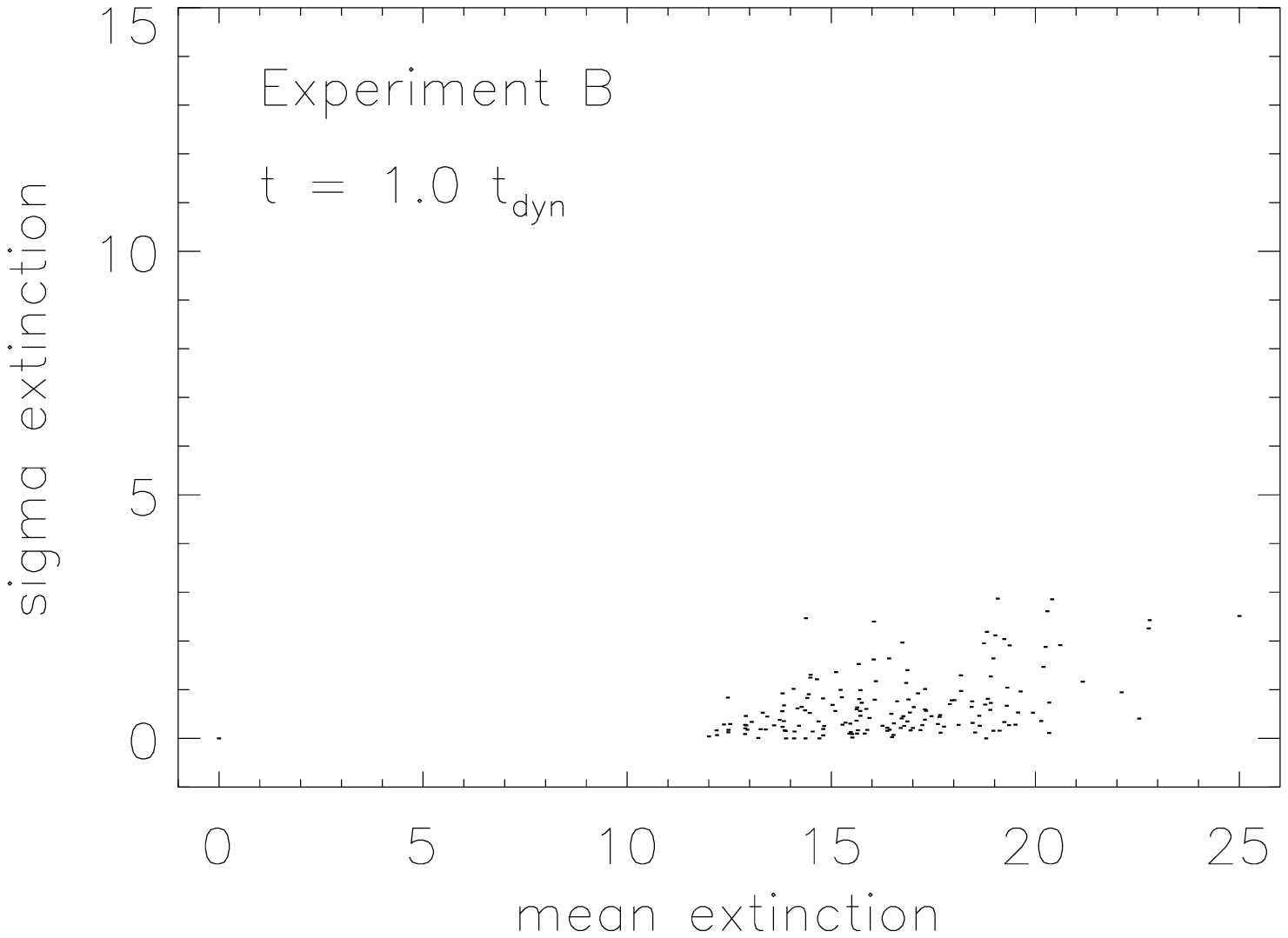}
    }
  \end{minipage}
\caption[]{Relation between the dispersion of the extinction and the mean
extinction, on a regular grid superposed of the projection of the density field.
The value of the extinction are measured at random positions, simulating
the random position of a star behind the cloud.}
\end{center}
\end{figure}

\clearpage
\begin{figure}
\begin{center}
  \begin{minipage}{80mm}
    \centerline{
    \epsfxsize=80mm
    \epsfbox{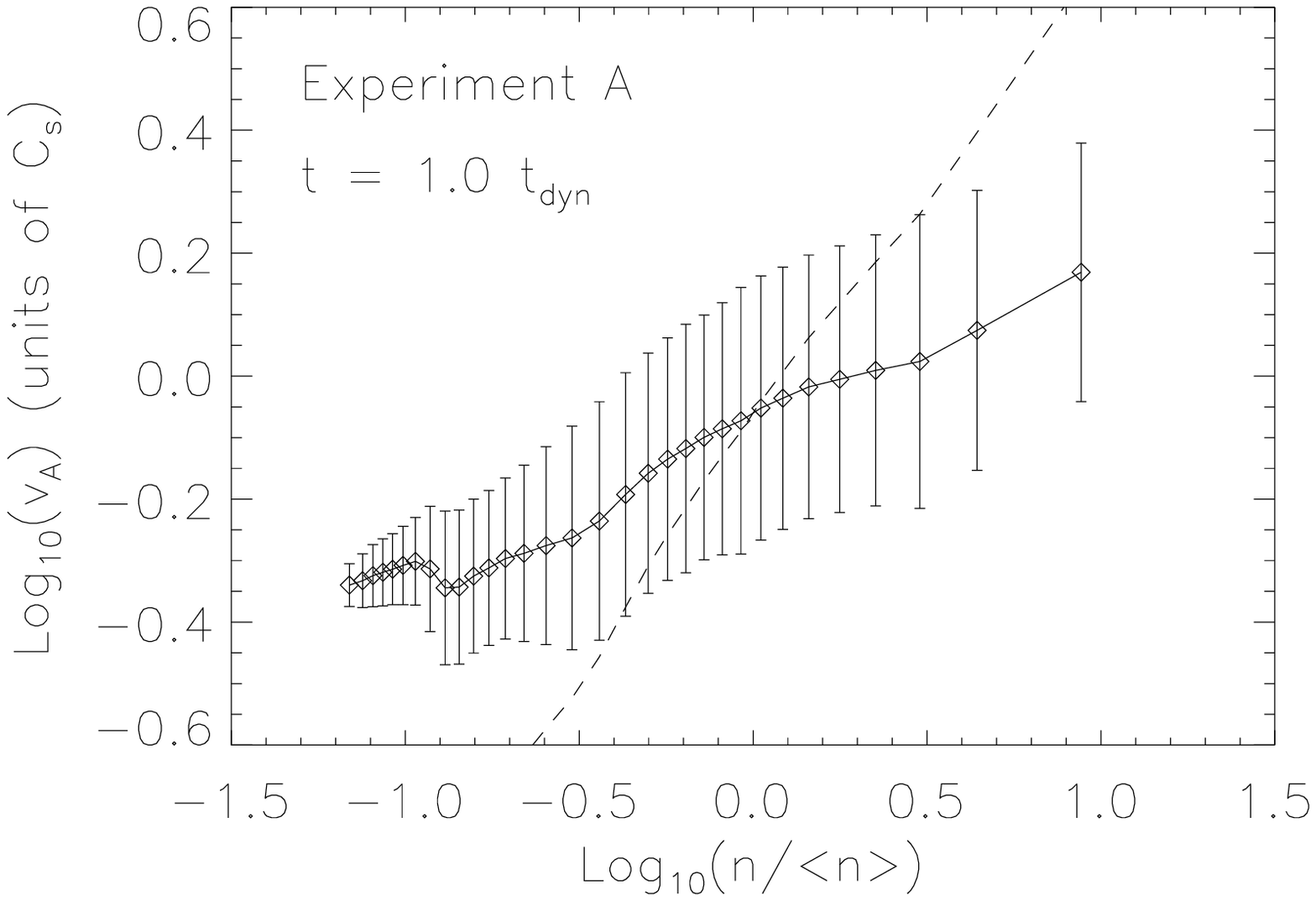}
    }
  \end{minipage}
  \begin{minipage}{80mm}
    \centerline{
    \epsfxsize=80mm
    \epsfbox{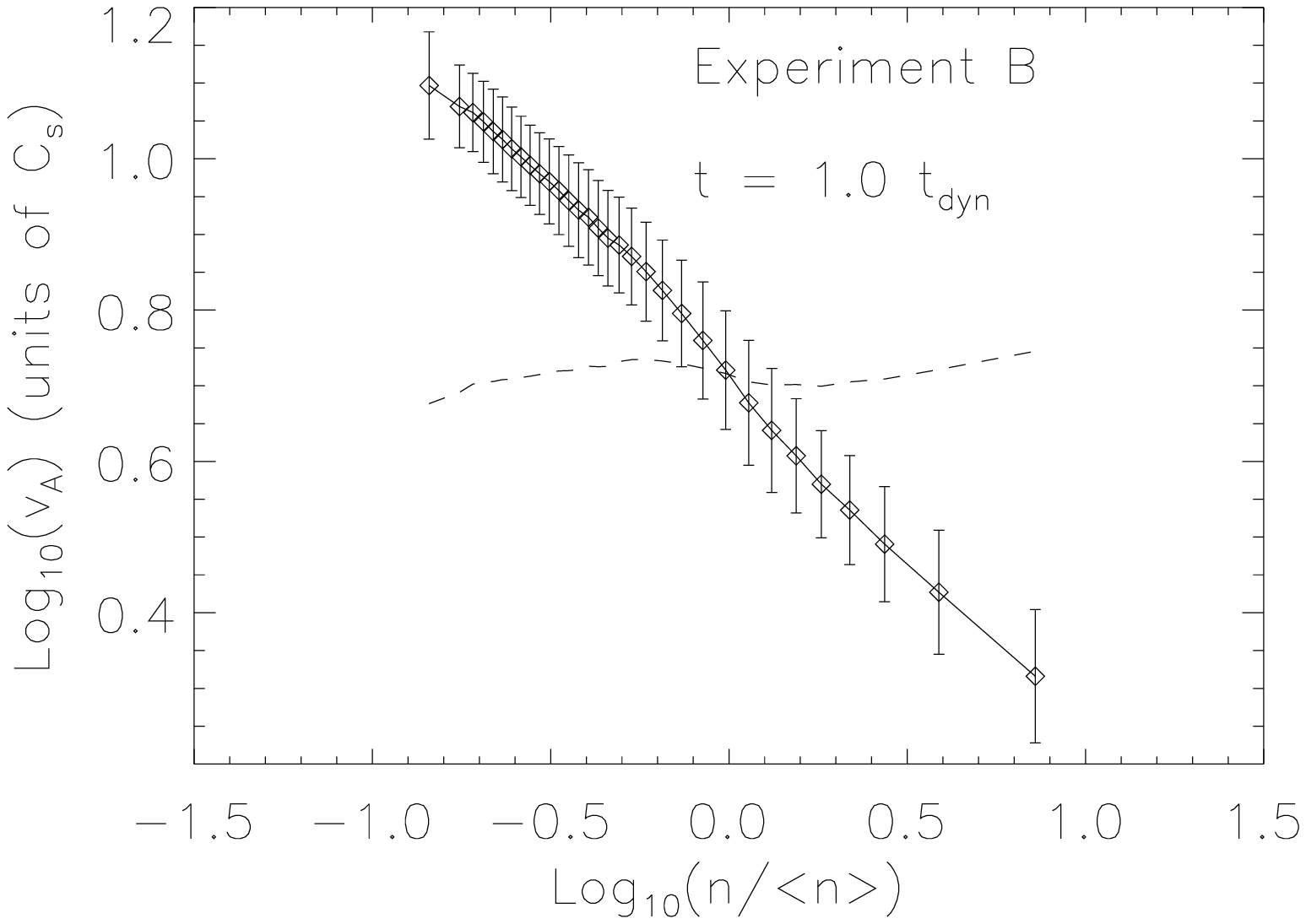}
    }
  \end{minipage}
\caption[]{The Alfv\'{e}n velocity, in units of the sound velocity, versus the gas
density. The dashed line is the $B-n$ relation, that is to say the Alfv\'{e}n 
velocity defined with the mean density, instead of with the local density. In 
Experiment A,
the magnetic pressure is on average larger than the gas pressure, in regions with 
gas density
larger than the mean.}
\end{center}
\end{figure}

\end{document}